\documentclass[reprint,superscriptaddress,amssymb,amsmath,aps,showpacs,10pt,longbibliography,prl,nofootinbib]{revtex4-2}
\usepackage{graphicx}
\usepackage{subfigure}
\usepackage{tabularx}

\usepackage{amsmath}
\usepackage{amssymb}
\usepackage{amsbsy}
\usepackage{wasysym}

\usepackage{titlesec}
\usepackage{bm}
\usepackage{comment}
\usepackage{verbatim}

\usepackage{dcolumn}

\usepackage{color}
\definecolor{darkblue}{rgb}{0,0.02,0.45}
\usepackage[colorlinks,bookmarks=false,citecolor=darkblue,linkcolor=red,urlcolor=blue]{hyperref}

\newcommand{\kv}{\mathbf k}

\newcommand{\cuf}{CuF$_2$}

\newcommand{\tbp}{\ensuremath{t_{b}^{+}}}
\newcommand{\tbcp}{\ensuremath{t_{bc}^{+}}}
\newcommand{\tbm}{\ensuremath{t_{b}^{-}}}
\newcommand{\tbcm}{\ensuremath{t_{bc}^{-}}}
\newcommand{\jbp}{\ensuremath{J_{b}^{+}}}
\newcommand{\jbcp}{\ensuremath{J_{bc}^{+}}}
\newcommand{\jbm}{\ensuremath{J_{b}^{-}}}
\newcommand{\jbcm}{\ensuremath{J_{bc}^{-}}}

\def\be{\begin{equation}}
\def\ee{\end{equation}}
\def\bea{\begin{eqnarray}}
\def\eea{\end{eqnarray}}

\def\vec{\mathbf}
\def\bs{\boldsymbol}
\def\mc{\mathcal}

\makeatletter
\newenvironment{suppbib}[1]{%
  \vspace{1.5em} 
  \noindent\rule{\columnwidth}{0.5pt} 
  \vspace{0.5em} 
  
  \small 
  
  \@ifundefined{c@NAT@ctr}{}{\setcounter{NAT@ctr}{0}}
  \setcounter{enumiv}{0}
  
  \list{\@biblabel{\@arabic\c@enumiv}}%
       {\settowidth\labelwidth{\@biblabel{#1}}%
        \leftmargin\labelwidth
        \advance\leftmargin\labelsep
        \usecounter{enumiv}%
        \let\p@enumiv\@empty
        \renewcommand\theenumiv{\@arabic\c@enumiv}%
        \setlength{\itemsep}{0pt}%
        \setlength{\parsep}{0pt}%
        \setlength{\parskip}{0pt}%
       }%
  \sloppy\clubpenalty4000\widowpenalty4000\sfcode`\.\@m
}{%
  \endlist
}
\makeatother

\begin{document}

\title{Magnon band splitting without altermagnetism in CuF$_2$}

\author{Ioannis Rousochatzakis}
\thanks{These two authors contributed equally to this work}
\email{I.Rousochatzakis@lboro.ac.uk}
\affiliation{Department of Physics, Loughborough University, Loughborough LE11 3TU, United Kingdom}

\author{Oleg Janson}
\thanks{These two authors contributed equally to this work}
\email{o.janson@ifw-dresden.de}
\affiliation{Institute for Theoretical Solid State Physics, Leibniz IFW Dresden, 01069 Dresden, Germany}

\author{Alexander A. Tsirlin}
\email{altsirlin@gmail.com}
\affiliation{Felix Bloch Institute for Solid-State Physics, University of Leipzig, 04103 Leipzig, Germany}

\begin{abstract}
Conclusive identification of an altermagnetic state requires going beyond mere symmetry arguments. We illustrate this in a combined computational and experimental study of the rutile-like material CuF$_2$, which is on the list of predicted altermagnets. Using \textit{ab initio} and linear spin-wave calculations supplied by magnetization measurements, we show that CuF$_2$ in its experimental monoclinic structure can be described by a spin-$\frac12$ model of weakly coupled square-lattice layers with the in-plane coupling $J_1\simeq 115$\,K and two synergistic antiferromagnetic interplane couplings amounting to 4\% and 8\% of $J_1$, respectively. Driven by long-range superexchange, these interlayer couplings are oblique to the square planes, resulting in the unit-cell doubling in the magnetically ordered state, thus effectively suppressing any altermagnetic band splitting.  Concurrently, we identify unusually strong Dzyaloshinskii-Moriya interactions, $|\mathbf D|/J_1\simeq 0.3$, that produce spin canting and, together with order-by-disorder effect, pin the N\'eel vector to the crystallographic $b$-axis. Additionally, DM anisotropy promotes magnon band splitting, but these bands remain non-chiral. Our results highlight the importance of relativistic effects even in $3d$ magnets with altermagnetic symmetries.
\end{abstract}

\maketitle

\section{INTRODUCTION}

Recently, altermagnetism emerged as a new concept in solid-state physics~\cite{smejkal2022}. Its central idea is that certain structural symmetries lead to the lifting of Kramers degeneracy in fully compensated magnetic structures~\cite{jungwirth2026}. In a metallic altermagnet, this would give rise to the anomalous Hall and spin Hall conductivities in the absence of a net magnetic moment. A counterpart effect in insulating altermagnets is the splitting of double-degenerate magnon bands into chiral, left- and right-polarized magnons~\cite{bai2024}. These band-splitting-driven effects lead to various functionalities and potential technological applications~\cite{song2025}. 

An initial identification of altermagnetic candidates is mainly based on their structural symmetry. The rutile structure is especially suitable in this regard because it can accommodate various transition-metal atoms as magnetic ions and contains the non-symmorphic $4_2$ symmetry operation connecting magnetic sublattices. The experimental situation turned out to be more involved, though, because the metallic rutile-type compound RuO$_2$ predicted to be an altermagnet~\cite{smejkal2022}, turned out to be nonmagnetic, at least in its bulk form~\cite{tsirlin2026}. Insulating $3d$ difluorides with the rutile structure are robust magnets and show a somewhat better agreement with theory. The anticipated splitting of the magnon bands and the chiral nature of the magnons could be confirmed experimentally in MnF$_2$~\cite{faure2025} and FeF$_2$~\cite{sears2026}, but the magnitude of this splitting is minute~\cite{morano2025}, suggesting dipolar couplings~\cite{faure2025,sears2026} or even magnon-phonon coupling~\cite{lovesey2026} instead of altermagnetic exchange terms as its primary origin. Given the weakness of dipolar couplings in $3d$ antiferromagnets and the sensitivity of dipolar energies to the sample shape, this renders magnon band splitting a minute effect of little practical relevance, at least in the rutile-type compounds studied to date.

Here, we address another rutile-type $3d$ difluoride, CuF$_2$, which is also on the initial list of altermagnetic candidates~\cite{smejkal2022}. Early studies of CuF$_2$ showed the \mbox{spin-$\frac12$} magnetism anticipated for Cu$^{2+}$ and the antiferromagnetic order below $T_N\!=\!71$\,K~\cite{boo1979,bastow1981} with a possible weak ferromagnetic component inferred from the small remnant magnetization~\cite{cuf2:joenk65}. Quantum chemical calculations established Cu--F--Cu superexchange as the primary origin of antiferromagnetism~\cite{cuf2:reinhardt99,cuf2:reinhardt00}, whereas magnetoelastic effect probed by high-resolution neutron diffraction~\cite{cuf2:chatterji11} revealed short-range antiferromagnetic order setting in around 150\,K, well above $T_N$, indicative of low-dimensional magnetism.

In the context of altermagnetism, two crucial experimental aspects of CuF$_2$ are its monoclinic symmetry due to a large distortion of the rutile structure~\cite{aramburu2019}, and the $\kv\!=\!(\frac12,0,0)$ nature of magnetic order pinpointed by neutron diffraction~\cite{cuf2:fischer74}. The latter observation unequivocally excludes altermagnetism because the doubling of the magnetic unit cell should restore the double degeneracy of the bands even for the altermagnetic symmetry of the parent crystal structure. Recent theoretical work on altermagnetism of CuF$_2$~\cite{muskan2026,ho2026,cuf2:leon26} mainly concentrates on the hypothetical scenario of the $\kv\!=\!0$ magnetic order inferred from the other $3d$ difluorides, although such an order was never seen in CuF$_2$ experimentally, nor can it be stabilized under practical conditions, as we demonstrate below. 

\begin{figure*}[!ht]
 \includegraphics{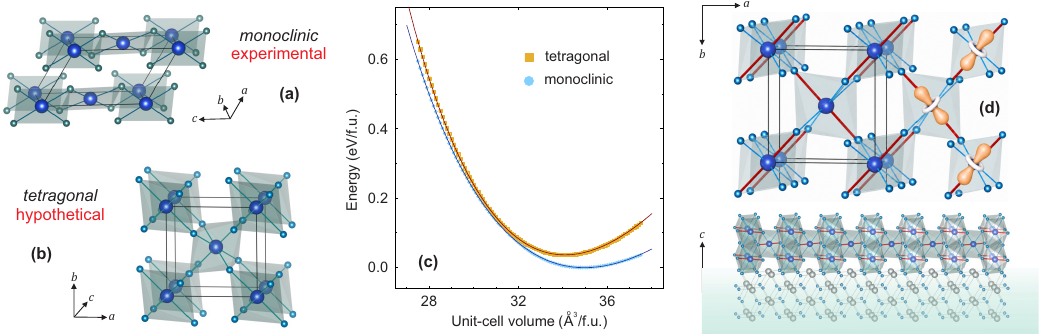}
 \caption{\label{fig:structure}
 \textbf{Crystal structure of CuF$_2$}. (a) Experimental monoclinic form with layers of the CuF$_4$ plaquettes. (b) Hypothetical tetragonal form with the ideal rutile structure. (c) Energy vs. volume calculated for both forms (minimum energy of the monoclinic form is chosen as zero energy). The lines show fits to the equation of state. (d) Stabilization of the tetragonal CuF$_2$ on the MgF$_2$ substrate. Red lines indicate shorter Cu--F bonds resulting in a staggered configuration of the $d_{3z^2-r^2}$ orbitals. VESTA software~\cite{vesta} was used for crystal structure visualization. }
\end{figure*}

In the present work, we then follow a different, experimentally-driven approach and analyze CuF$_2$ in its real, distorted rutile structure. We identify the origin of the $\kv\!\neq\!0$ magnetic order and also demonstrate the relevance of anisotropic, Dzyaloshinskii-Moriya (DM) exchange to the physics of this material. The importance of the DM interactions sets interesting parallels between CuF$_2$ and parent compounds of high-$T_c$ cuprates, as both share a sizable spin canting. Including the DM interactions is also essential for reproducing thermodynamic properties of CuF$_2$. Intriguingly, the spin-wave spectra calculated for the realistic spin model of CuF$_2$ show a splitting of magnon bands, the hallmark of altermagnetism, but this splitting has non-altermagnetic (relativistic) origin, and individual magnon bands are non-chiral, in contrast to the altermagnetic scenario.

\section{RESULTS}
\subsection{Crystal structure}
\label{sec:structure}

Our experimental XRD data confirm the monoclinic structure of CuF$_2$ (Fig.~\ref{fig:structure}a). The monoclinic distortion is mainly related to the deformation of CuF$_6$ octahedra into CuF$_4$ plaquette units, which are typical for Cu$^{2+}$ as a Jahn-Teller ion. Although the idealized rutile-type tetragonal structure of altermagnetic CuF$_2$ (Fig.~\ref{fig:structure}b) clearly does not occur at ambient conditions, one may ask whether it could be stabilized by tailoring the material with pressure or strain.

To this end, we calculated the total energies of the tetragonal (idealized) and monoclinic (experimental) CuF$_2$ for a series of fixed volumes upon relaxing unit-cell shape and atomic positions. The resulting energy vs. volume curves (Fig.~\ref{fig:structure}c) were fitted with the Birch-Murnaghan equation of state to obtain the equilibrium volume $V_0=34.99$\,\r A$^3$/f.u., bulk modulus $B_0=76.9$\,GPa, and its pressure derivative $B_0'=4.41$ for the monoclinic structure as well as $V_0=34.10$\,\r A$^3$/f.u., $B_0=104.1$\,GPa, and $B_0'=4.50$ for the tetragonal structure. The monoclinic structure is 37\,meV/f.u. more stable in equilibrium. This energy difference decreases upon compression, but the monoclinic structure remains lower in energy at all pressures. It means that pressure does not invert the Jahn-Teller distortion of Cu$^{2+}$, and CuF$_2$ is unlikely to become tetragonal under hydrostatic compression.

More favorable conditions for stabilizing the tetragonal structure could be potentially created by depositing a CuF$_2$ thin film on a tetragonal or cubic substrate. We thus considered several substrates that span the possible range of the in-plane lattice parameters: MgO ($a=4.213$\,\r A), MgF$_2$ ($a=4.62$\,\r A), and Y-doped ZrO$_2$ ($a=5.13$\,\r A). Structure relaxations were performed with fixed $a=b$ while varying the $c$-parameter and atomic positions. The relaxed $c$-value systematically decreased with increasing $a$. Additionally, the CuF$_6$ octahedra demonstrated a characteristic deformation with two shorter and four longer Cu--F bonds that lead to the half-filling of the $d_{3z^2-r^2}$ orbital where the local $z$-axis is directed along the shorter Cu--F bonds. The direction of this local axis alternates between the neighboring Cu atoms, thus resulting in a staggered orbital order (Fig.~\ref{fig:structure}d) and a strong ferromagnetic interaction between nearest neighbors. For example, in CuF$_2$ on MgF$_2$ (001) the ferromagnetic spin configuration is stabilized by 60.4\,meV/f.u.\ compared to the nearest-neighbor antiferromagnetic one. 

We thus conclude that stabilizing the idealized tetragonal structure of CuF$_2$ by strain would simultaneously induce ferromagnetism in this material. Spin-compensated order in CuF$_2$ under experimentally relevant conditions is possible in the currently known monoclinic phase only. We will now analyze the microscopic aspects of this monoclinic phase and uncover the nature of its magnetic order being antiferromagnetic rather than altermagnetic. In the following, we will use the experimental lattice parameters of monoclinic CuF$_2$. Note also that the standard $P2_1/c$ setting of the monoclinic space group 14 is chosen throughout this study.

\subsection{Electronic structure}

\begin{figure}[tb]
\includegraphics[width=8.6cm]{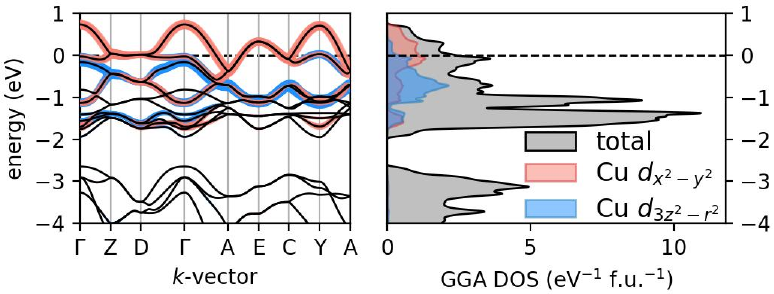}
\caption{\label{fig:dosband} \textbf{ Band structure of CuF$_2$}.  Nonmagnetic GGA band structure (left) and the density of states (right) of CuF$_2$. Contributions of Cu $d_{x^2-y^2}$ and $d_{3z^2-r^2}$ are highlighted. Notation of high-symmetry $k$-vectors: Z=$(0,-\frac12,0)$, D=$(0,-\frac12,\frac12)$, A=$(\frac12,0,-\frac12)$, E=$(\frac12,-\frac12,-\frac12)$, C=$(\frac12,-\frac12,0)$, Y=$(\frac12,0,0)$.}
\end{figure}

As expected for a Cu$^{2+}$-containing compound, the nonmagnetic GGA calculation for \cuf\ yields a spurious metallic solution. However, the sizable ionicity of Cu--F bonds splits the valence band into a higher-lying $d$-dominated and a lower-lying $p$-dominated manifold. The former comprises ten bands (two Cu sites per cell with five $d$ orbitals each), two thereof cross the Fermi level (Fig.~\ref{fig:dosband}). By inspecting the orbital character of these states, we find that the highest-lying bands are dominated by $d_{x^2-y^2}$, yet in the energy window between $-1$ eV and the Fermi level they are intertwined with $d_{3z^2-r^2}$ states. Due to this overlap, a construction of an effective one-orbital ($d_{x^2-y^2}$) model becomes very challenging, and the Fourier transform of the resulting Wannier Hamiltonian significantly deviates from the GGA bands. Despite this shortcoming, the resulting model correctly captures the leading intra- and interplane couplings; its completeness is \textsl{a posteriori} confirmed by very low error bars on the exchange integrals obtained from GGA+$U$ total energies (see below).

\begin{figure}[tb]
\includegraphics[width=8.6cm]{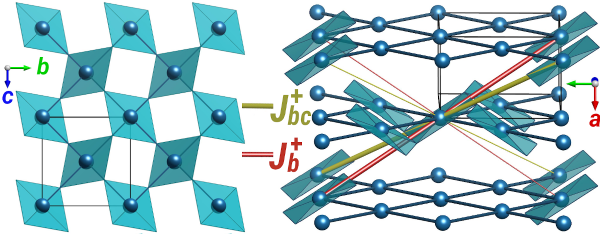}
\caption{\label{fig:model} \textbf{Crystal structure and minimal spin model of CuF$_2$}. Left: CuF$_4$ squares (cyan) forming a square lattice in the $bc$ plane. Projections of interplane interactions \jbp\ and \jbcp\ are shown, respectively, as red and yellow cylinders. Right: connectivity of magnetic planes. For clarity, only bonds connecting the central Cu atom are shown. Relevant interplane interactions \jbp\ and \jbcp\ are shown as thick cylinders, their altermagnetic counterparts \jbm\ and \jbcm\ as thin lines. Note the different mutual orientation of CuF$_4$ plaquettes connected by (\jbp, \jbcp) and (\jbm, \jbcm).}
\end{figure}

The Wannier Hamiltonian contains four sizable terms. In the square-lattice planes (Fig.~\ref{fig:model}, left), these are the nearest-neighbor hopping $t_1\!=\!260$\,meV and the second-neighbor hopping operating along one of the diagonals $t_2'\!=\!56$\,meV. The latter is sizable, because it connects coplanar CuF$_2$ units; along the other diagonal the neighboring units are tilted, making the respective hopping negligible. Two further relevant terms couple the magnetic planes together: \tbp$=\!31$\,meV and \tbcp$=\!46$\,meV. The former is again underpinned by the mutual alignment of CuF$_4$ units, while the latter is assisted by efficient Cu-F..F-Cu connections. In our notation, subscripts $b$ and $c$ denote the corresponding lattice translations, while the superscript $+$ indicates that both couplings have an (altermagnetic) counterpart, with the same Cu..Cu distance, but a different crystalline environment: CuF$_4$ units connected by \tbm\ are noncoplanar, while \tbcm\ lacks the F..F bridge in between the Cu atoms. As a result, \tbm\ and \tbcm\ are negligibly small (below 3 meV). This disparity is illustrated in Fig.~\ref{fig:model}, right.

Due to its one-orbital nature, the effective model does not fully describe the magnetism of CuF$_2$, yet the standard 
perturbative expression $J\propto{}t^2$ allows us to draw several qualitative considerations. First, \cuf\ is a quasi-2D square-lattice antiferromagnet with the sizable $J_1$ exchange -- the dominance of $t_1$ excludes other scenarios. Second, the second-neighbor exchange $J_2'$ is more than an order of magnitude smaller, $(t_2'/t_1)^2=0.046$; it is expected to give rise to weak magnetic frustration only. Third, the magnetic planes are connected by two exchanges in a nonfrustrated manner. Crucially, both interlayer couplings connect CuF$_4$ units of the same spatial orientation. Hence, in the anticipated magnetically ordered state, these units will have opposite spin polarization, and the overall altermagnetic spin splitting in the bulk will vanish.

\subsection{Calculation of exchange couplings} 
Next, we estimate the magnetic exchanges $J_1$, $J_2'$, \jbp\, and \jbcp\ using spin-polarized GGA+$U$ total-energy calculations. The dominant exchange $J_1$ amounts to $154\pm20$\,K for $U_d\!=\!8\mp1$ eV. The $J_2'$, \jbp\, and \jbcp\ exchanges show a similar dependence on the $U_d$ parameter and amount, respectively, to 5, 4, and 8\% of $J_1$. To assess the numerical accuracy of these values, we fixed $U_d\!=\!8$ eV and performed calculations for a larger supercell that allowed us to distinguish between \jbcp\ and \jbcm; the latter was found to be weakly ferromagnetic \jbcm$=\!-0.7$\,K. Importantly, estimates obtained from total energies of two different supercells are in excellent agreement, justifying our restriction to only four exchanges: $J_1$, $J_2'$, \jbp\, and \jbcp.

Our analysis of the nonmagnetic band structure of CuF$_2$ revealed a sizable mixing of the magnetically active $x^2\!-\!y^2$ orbital with the closely lying $3z^2\!-\!r^2$ orbital. Such a mixing is generally favorable for exchange anisotropy. To address this possibility, we computed the full bilinear exchange matrix between the two NN sites $(0,0,0)$ and $(0,\frac12,\frac12)$. In this way, we found a strong Dzyaloshinskii-Moriya (DM) anisotropy $|\vec{D}|/J_1\!\simeq\!0.3$, with the DM vectors perpendicular to respective Cu..Cu bonds and slightly tilted out of the plane of the square lattice. While the presence of the DM component can be generally expected due to the absence of inversion symmetry connecting the neighboring CuF$_4$ units, its large magnitude is rather unusual. Since $|\vec{D}|$ is roughly 4 to 8 times stronger than any of longer-range isotropic exchanges ($J_2'$, \jbp\, and \jbcp), we expect it to play a significant role in the magnetism of \cuf.

\subsection{Magnetic susceptibility fit} 
\begin{figure}[tb]
\includegraphics[width=8.6cm]{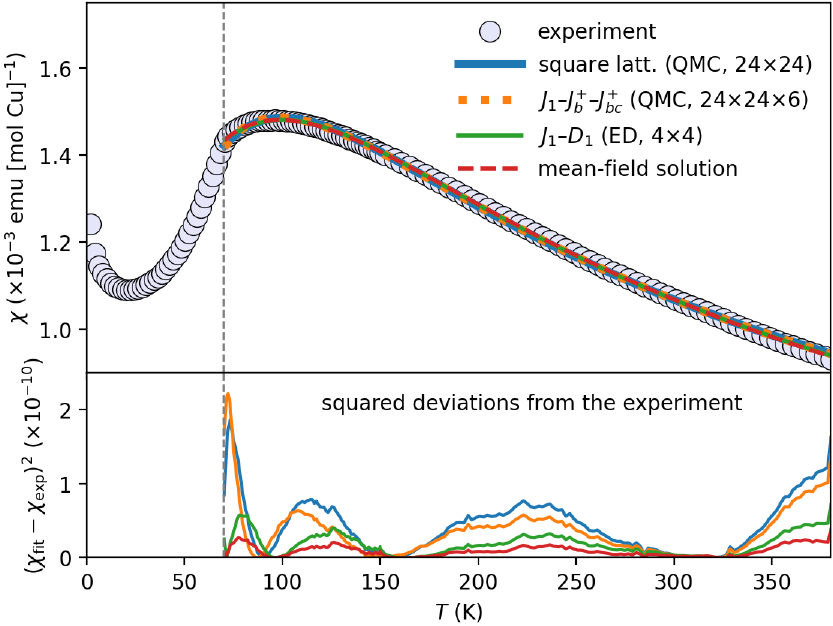}
\caption{\label{fig:fit_chi} \textbf{Magnetic susceptibility of \cuf}. Fits (lines) to the experimental magnetic susceptibility of \cuf\ (circles). The refined parameters $J_1$, $g$, and $\chi_0$ are provided  in Table~\ref{tab:chi_fit_params}. Bottom panel shows squared deviations. The vertical dashed line depicts the long-range magnetic
ordering temperature.}
\end{figure}

To further refine the parameters of the spin model, we turn to model simulations. Since the magnetism of \cuf\ is dominated by $J_1$, a natural choice for the minimal model is the $S\!=\!\frac12$ Heisenberg square lattice. In the paramagnetic regime, the QMC-simulated curve generally agrees with the experimental data (Fig.~\ref{fig:fit_chi}, top), yet deviations are clearly visible, especially if we inspect the squared differences (Fig.~\ref{fig:fit_chi}, bottom). Moreover, the fitted value of the $g$-factor (1.940, see Table~\ref{tab:chi_fit_params}) is too small for Cu$^{2+}$ that should normally show $g\!>\!2.0$. One possible source of deviations is the neglect of interplane couplings. To check this, we supplement the model with \jbp\ and \jbcp\ (the weak $J_2'$ is omitted to avoid the sign problem) and perform QMC simulations for the resulting 3D model. The agreement is somewhat improved, yet the $g$-factor of 1.975 is still too small.

\begin{table}[tb]
\caption{\label{tab:chi_fit_params}Parameters obtained from a least-squares fit
to the experimental $\chi(T)$ data by using four different spin models.}
\begin{ruledtabular}
\begin{tabular}{rddr}
model (method)         & \multicolumn{1}{c}{$J_1$ (K)} & \multicolumn{1}{c}{$g$-factor} & \multicolumn{1}{c}{$\chi_0 (\frac{\text{emu}}{\text{mol Cu}})$} \\ \hline
$J_1\equiv$ square lattice (QMC) & 106.3 & 1.940 & $2.4\times10^{-4}$ \\
$J_1$--\jbp--\jbcp (QMC)         & 106.1 & 1.975 & $2.2\times10^{-4}$ \\
$J_1$--DM (ED)                & 114.9 & 2.022 & $1.9\times10^{-4}$ \\
full (mean field)                & 120.4 & 2.115 & $1.5\times10^{-4}$ \\
\end{tabular}
\end{ruledtabular}
\end{table}

Since interplane exchanges apparently play only a minor role in the $\chi(T)$ behavior, further improvement of the fit is possible only if the exchange anisotropy is taken into account. However, such models do not admit the use of efficient numerical techniques for a computation of thermodynamic properties. In practice, the non-commutativity of $\hat{S}^z$ and $\hat{H}$ leaves us with the only option -- performing a full diagonalization of the Hamiltonian, which is possible only for small lattices due to the exponential growth of the matrix size. Here, we consider an idealized $J_1$-DM model with DM vectors pointing along the diagonals of the square lattice, i.e., we neglect small monoclinic distortions and the out-of-plane tilting of the DM vectors. We compute the magnetic susceptibility on $4\times4$ finite lattices for three different spatial directions of magnetic field: i) in-plane along half of the DM vectors, ii) in-plane in-between the DM vectors (at 45$^{\circ}$ to the first direction), and iii) perpendicular to the planes. The final magnetic susceptibility used for fitting is obtained by averaging over these three directions. The fit quality (Fig.~\ref{fig:fit_chi}) improves considerably, and the anticipated $g$-value above 2.0 is readily obtained.

Finally, we consider the complete model, with all four Heisenberg exchanges and the DM anisotropy of the leading term. Profiting from the dominance of $J_1$, we can use the square-lattice model as a starting point, and account for DM, $J_2'$, \jbp\ and \jbcp\ terms in a mean-field approximation (see details in the Supplemental material (SM), Sec.~V). Setting $J_2'$, \jbp\ and \jbcp\ equal to, respectively, 4, 4, and 8\% of $J_1$, we obtain a theoretical curve, which gives the best agreement with the experiment (Fig.~\ref{fig:fit_chi}). We also note that the $g$-factor of 2.115 falls into the range expected for Cu$^{2+}$ and the fitted $J_1$ becomes closer to the DFT+$U$ estimates (Table~\ref{tab:chi_fit_params}).

\begin{figure*}
 \includegraphics{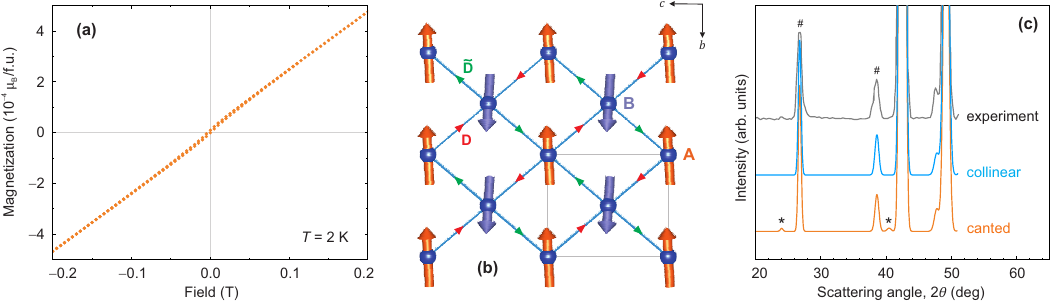}
 \caption{\textbf{Magnetization process and magnetic ground state of CuF$_2$}. (a) Field-dependent magnetization measurement at 2\,K shows the absence of weak ferromagnetism. (b) Canted in-plane magnetic structure of CuF$_2$ expected from our microscopic magnetic model. Weak ferromagnetic moment produced by spin canting is compensated by the opposite moment in the adjacent layer. Small arrows on the bonds show the site ordering convention $(i\to j)$ for the corresponding DM vector ${\bf D}_{ij}$. (c) Experimental neutron diffraction pattern~\cite{cuf2:fischer74} (wavelength 2.347\,\r A) along with the simulated patterns for the collinear magnetic structure from Ref.~\cite{cuf2:fischer74} and for the canted structure proposed in our work. Hash symbols indicate two most intense magnetic Bragg peaks, whereas asterisks label weak additional peaks that arise from spin canting. The $b$-component of the magnetic moment is set to 0.68\,$\mu_B$ per Ref.~\cite{cuf2:fischer74}.}
 \label{fig:gs} 
\end{figure*}

\subsection{Magnetic ground state} 
Magnetic structure of CuF$_2$ is usually described as fully compensated and collinear. However, weak ferromagnetism with a small uncompensated moment of $2\times 10^{-4}$\,$\mu_B$/f.u. was also reported~\cite{cuf2:joenk65}. The presence of weak ferromagnetism is diagnostic for the ground state because it would imply an incomplete compensation between the adjacent layers and the presence of a $\mathbf k=0$ ordering component. We thus performed field-dependent magnetization measurements at 2\,K. Our data (Fig.~\ref{fig:gs}a) do not confirm weak ferromagnetism and set the upper boundary of less than $1\times 10^{-5}$\,$\mu_B$/f.u. on the remnant magnetization. The magnetic ground state of CuF$_2$ is, therefore, fully compensated. 

A fully compensated ground state is not necessarily collinear. A weak ferromagnetic moment may form within each layer as a result of canting between the sublattices A and B (Fig.~\ref{fig:gs}b), but the moments of two adjacent layers compensate each other because of the $\mathbf k=(\frac12,0,0)$ periodicity of the magnetic structure. 

Let us now discuss the nature of the intralayer order predicted by our microscopic magnetic model. 
A detailed analysis is given in the SM (Secs.~III-IV) and the key results can be summarized as follows: 
i) The spins lie in the plane perpendicular to the vector ${\bf d}\!=\!(D_x,0,D_z)$, where $D_x\!=\!15.5$\,K and $D_z\!=\!26.6$\,K are the $x$ and $z$ components of ${\bf D}$ (where, in our convention, $(x,y,z)$ is the $(a^\ast, b,c)$ frame).    
ii) The angle $\phi$ that quantifies the deviation from the fully antiparallel arrangement of NN spins is determined by $\tan\phi\!=\!d/J_1$, where $d\!=\!|{\bf d}|$, and amounts to approximately $14^\circ$ for our coupling parameters. 
iii) At the  classical level, a global rotation around ${\bf d}$ does not change the energy. This accidental degeneracy is lifted by quantum fluctuations (order by disorder effect). According to linear spin-wave theory (see SM, Sec.~IV), the preferred direction of the N\'eel order parameter is the $b$ axis, and that of the weak FM moment is along the vector ${\bf d}_\perp\!=\!(-D_z,0,D_x)$, see also Fig.~\ref{fig:gs}b.

This structure should be contrasted to the one reported experimentally in Ref.~\cite{cuf2:fischer74}, wherein spins are tilted by $(21\pm6)^\circ$ away from the $b$ axis, but remain strictly collinear (this structure corresponds to the Shubnikov group $P_a2_1/c$). Such a collinear state would be in fact unexpected in the presence of the DM couplings, and would be unlikely from the symmetry perspective too. Indeed, as discussed in the SM, the components $(L_x,F_y,L_z)$ and $(F_x,L_y,F_z)$ transform according to the irreps $\Gamma_1$ and $\Gamma_3$ of the crystallographic group, respectively (where ${\bf F}$ and ${\bf L}$ are the in-plane ferromagnetic moment and N\'eel vector, respectively). Generically, a second-order phase transition at $T_N$ that gives rise to a nonzero $|{\bf L}|$ should follow one of these irreps. The collinear structure proposed in Ref.~\cite{cuf2:fischer74} requires a combination of the $\Gamma_1$ and $\Gamma_3$ irreps, and thus does not satisfy this condition. By contrast, the canted magnetic structure that resulted from our analysis (Fig.~\ref{fig:gs}b) follows the $\Gamma_3$ irrep and is thus allowed by symmetry.

We have further used Mag2Pol~\cite{qureshi2019} to simulate the neutron diffraction pattern of CuF$_2$ assuming our canted model with the canting angle of $14^{\circ}$ as well as the collinear model from Ref.~\cite{cuf2:fischer74}. Two intense magnetic Bragg peaks are the same between the two models, as they are determined by the primary $b$-component of the magnetic moment (Fig.~\ref{fig:gs}c). The canting gives rise to two additional, very weak magnetic Bragg peaks that are likely below the sensitivity limit of the available diffraction experiment. We thus conclude that our magnetic model is consistent with the experimental neutron diffraction data for CuF$_2$, but further experiments with the improved sensitivity would be highly desirable in order to detect the weak magnetic Bragg peaks associated with the spin canting.

\begin{figure}[!t]
\includegraphics[width=0.49\textwidth,angle=0,clip=true,trim=0 0 0 0]{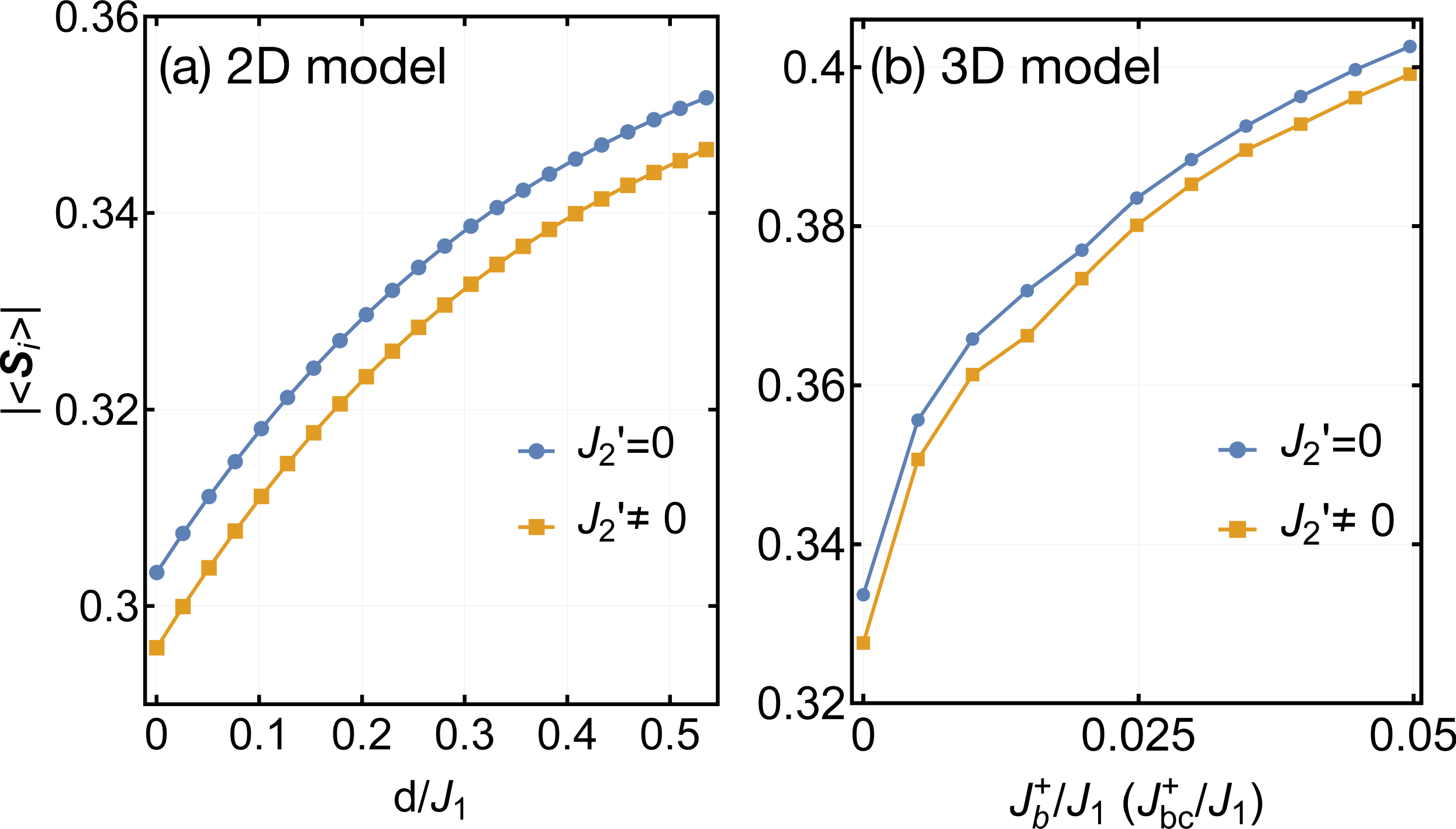}
\caption{\textbf{Spin length}. Impact of DM anisotropy and interlayer couplings on the local spin length, as calculated from linear spin wave theory. (a) Spin length vs $d/J_1$ for a single layer of CuF$_2$, with and without $J_2'$. (b) Spin length vs $J_b^+/J_1$ (keeping $J_{bc}^+\!=\!J_b^+$ and $d\!\neq\!0$ fixed) for the 3D model, with and without $J_2'$.}\label{fig:SpinLength}
\end{figure}

Finally, let us analyze the value of the local magnetic moment, which is found experimentally from neutron diffraction to be approximately $0.73\mu_B$~\cite{cuf2:fischer74}. Assuming $g\!\simeq\!2.115$, this gives a local spin length of $\langle S_i\rangle\!\simeq\!0.345$, which is higher than the value of $0.303$ predicted for the  spin-$\frac12$ NN Heisenberg model on the square lattice, based on  linear spin-wave theory~\cite{Anderson1952}. This difference can be attributed to the DM and the interlayer interactions, which both have the tendency to reduce the QM fluctuations. By contrast, the $J_2'$ coupling frustrates the dominant in-plane N\'eel order and is therefore expected to enhance the QM fluctuations. 

Figure~\ref{fig:SpinLength}(a) shows the value of the local spin length $\langle S_i\rangle$ as a function of $d/J_1$, as calculated from linear spin-wave theory for a single layer of CuF$_2$. The two curves show the values with and without the $J_2'$ coupling. As expected, the spin length is lower in the presence of $J_2'$. Furthermore, $\langle S_i\rangle$ increases monotonically with $d/J_1$, and reaches a value of $0.328$ for our \textit{ab initio} coupling parameters ($d/J_1\!\simeq\!0.255$ and $J_2'/J_1\!\simeq\!0.05$). 

A further increase is anticipated due to the interlayer couplings. This can be seen in Fig.~\ref{fig:SpinLength}(b), which shows the behavior of $\langle S_i\rangle$ for the full 3D model, as we vary the ratio $J_b^+/J_1$ (keeping $J_{bc}^+\!=\!J_b^+$ fixed) from zero up to the {\it ab initio} value of $0.05$, and with DM interactions present (with $d/J_1\!\simeq\!0.255$). The two curves show again the spin length with and without $J_2'$. The spin length increases monotonically with the strength of the interlayer couplings, reaching $\langle S_i\rangle\!\simeq\!0.4$ at $J_b^+/J_1\!\simeq\!0.05$. This value exceeds the one found experimentally, suggesting that our {\it ab initio} calculations may overestimate $d$, $J_b^+$ and $J_{bc}^+$ and/or underestimate $J_2'$.


\subsection{Elementary excitations}
Having established the minimal microscopic model for the magnetic ground state of CuF$_2$, we next explore the magnon excitation spectrum based on linear spin-wave theory (see SM, Sec.~VI). We have considered both a single layer of CuF$_2$ and the full 3D structure, and have obtained analytical expressions for the corresponding magnon frequencies, see SM. Here, we shall focus on the 3D structure. 
We have also calculated the zero-temperature, dynamical spin structure factor component $\mc{S}_{\perp}({\bf q},\omega)$ measured in unpolarized neutron scattering experiments, as well as the `chiral structure factor' $\mc{C}({\bf q},\omega)$ measured in polarized neutron scattering (with polarization along ${\bf q}$) or by RIXS, which  reveals if the magnon modes are chiral~\cite{Weber2018,Jenni2022,Smejkal2023,Holbein2023,Liu2024,McClarty2025,faure2025,Biniskos2025,cuf2:leon26,ho2026}.

\begin{figure*}[!t]
\includegraphics[width=\textwidth,angle=0,clip=true,trim=0 0 0 0]{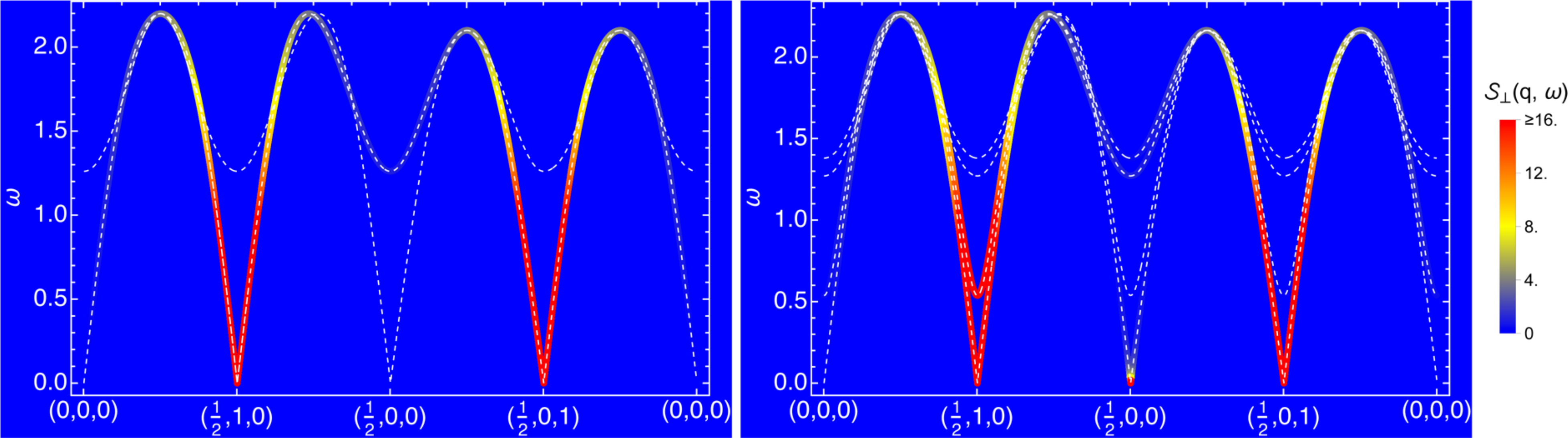}
\caption{\textbf{Elementary excitations}. Linear spin-wave theory results for CuF$_2$, described by the $J_1$-$J_2'$-$J_b^+$-$J_{bc}^+$-DM model  with $J_1\!=\!121$\,K, $J_2'\!=\!J_{bc}^+\!=\!J_{b}^+\!=\!6$\,K, and $d\!=\!0$ (panel a) or $d\!\neq\!0$ (panel b). Dashed lines show the magnon dispersions $\omega({\bf q})$ (in units of $J_1$) along special directions in momentum space (in r.l.u.). Colours show the dynamical spin structure factor component measured in unpolarized inelastic neutron scattering experiments, $\mc{S}_{\perp}({\bf q},\omega)\!=\!\int \frac{dt}{2\pi} e^{i \omega t} \langle \delta{\bf M}_{\perp}({\bf q}, t)\!\cdot\!\delta{\bf M}_{\perp}(-{\bf q}, 0)\rangle$, where $\delta{\bf M}_{\perp}({\bf q}, t)\!=\!\hat{\bf q}\!\times\!(\delta{\bf M}({\bf q},t)\!\times\!\hat{\bf q})$, $\hat{\bf q}\!=\!{\bf q}/|{\bf q}|$, and $\delta{\bf M}({\bf q},t)$ is the spatial Fourier transform of the magnetization fluctuations.}
\label{fig:3DSqw} 
\end{figure*}

CuF$_2$ features four magnon modes due to the doubling of the magnetic unit cell along the $a$ axis and the inequivalence between the A and B sites. The magnon dispersion and corresponding data for $\mc{S}_\perp({\bf q},\omega)$ are shown in Fig.~\ref{fig:3DSqw} for models with and without DM interactions. 
In the absence of the DM terms (panel a), we obtain two split modes, two-fold degenerate each, as expected in simple antiferromagnets. The origin of this degeneracy can be understood based on the fact that a sublattice spin rotation on the down-spin sites (B sites on even layers and A sites on odd layers) does not lower the translational symmetry of the spin Hamiltonian (which is larger than the actual one, since the Heisenberg couplings do not differentiate between the A and B sites). Finally, the $\mc{S}_\perp({\bf q},\omega)$ data show the expected Goldstone modes with large intensity at the magnetic Bragg peak positions $(\frac12,1,0)$ and $(\frac12,0,1)$.

Figure~\ref{fig:3DSqw}\,(b) shows the corresponding results with DM terms included. The main features can be summarized as follows.
First, the degeneracy of each of the branches of panel (a) is now lifted, giving rise to four visually distinct bands, as one would expect in an altermagnet. The splitting is largest in the region around the magnetic Bragg peaks. Importantly, this splitting is driven by the DM anisotropy (see analytical expression in the SM) and is not altermagnetic in origin. 
Second, the remaining zero-energy modes are artifacts of linear spin-wave theory, and  reflect the accidental classical ground state degeneracy mentioned above. These modes are expected to gap out by nonlinear spin-wave corrections. Third, $\mc{S}_\perp({\bf q},\omega)$ is large for the pseudo-Goldstone modes at the magnetic Bragg peak positions. In addition, compared to panel (a), a nonzero intensity now also appears for the low-energy mode at $(\frac12,0,0)$. This is associated with the fluctuations of the weak FM moment, which alternates from one layer to the next. 
Finally, we also note the asymmetry between the large intensity of the second mode at $(\frac12,1,0)$ vs the vanishing intensity of the second mode at $(\frac12,0,1)$. This asymmetry reflects the special role of the $b$-axis, which is the direction of the N\'eel vector, as selected by quantum fluctuations.

Finally, we discuss our results for the  chiral factor, defined as $\mc{C}({\bf q},\omega)\!=\!-i\int \frac{dt}{2\pi} e^{i \omega t} \langle \hat{\bf q}\cdot (\delta{\bf M}({\bf q},t) \!\times\! \delta{\bf M}(-{\bf q},0))\rangle$. As shown in the SM (Sec.~VI), for a single $bc$ layer of CuF$_2$, one of the two magnon modes acquires nonzero chirality in the presence of DM anisotropy. The sign of this chirality depends on the direction of the N\'eel vector ${\bf L}$, due to time-reversal. As a result, in the 3D system, the total chirality cancels out in the presence of DM anisotropy. 
This should be contrasted to the hypothetical altermagnetic  scenario (not realized in CuF$_2$) where the interlayer couplings are ferromagnetic and not antiferromagnetic. In this scenario, the two magnon modes acquire opposite chiralities. Hence, the chiral structure factor probed by polarized neutron scattering can be the direct experimental tool for distinguishing between altermagnetic and DM-induced magnon band splitting in MnF$_2$.

\section{DISCUSSION}
Our microscopic analysis backed by the experimental magnetic susceptibility data identifies CuF$_2$ as a quasi-2D square-lattice antiferromagnet. Its magnetism is distinct from that of rutile-type dioxides and difluorides, which are conventional 3D antiferromagnets. This difference is rooted in the Jahn-Teller effect of Cu$^{2+}$ stabilizing the $x^2-y^2$ orbital and ensuing monoclinic symmetry of the crystal structure. Our \textit{ab initio} results suggest the persistence of this monoclinic distortion also under hydrostatic pressure. Biaxial strain from a suitably chosen substrate may enforce tetragonal symmetry, but it concurrently stabilizes ferromagnetism  promoted by the magnetically active $3z^2-r^2$ orbital. We conclude that no tuning is expected to produce an altermagnet, because the orbital switch ($x^2-y^2 \rightarrow 3z^2-r^2$) stabilizing the rutile structure is inextricably coupled with a magnetic switch making the leading $J_1$ ferromagnetic. 

The close magnetic siblings of CuF$_2$ are spin-$\frac12$ square-lattice antiferromagnets, such as La$_2$CuO$_4$, the canonical material for studying the microscopic mechanism of anisotropic exchange in cuprates~\cite{coffey1990,coffey1991}. CuF$_2$ exhibits a similar spatial configuration of the DM vectors, yet with a much larger magnitude of $|{\bf D}|/J_1$. Microscopically, this enhanced DM anisotropy can be linked to the large twisting of the CuF$_4$ plaquettes (Cu--F--Cu angle of $132.3^{\circ}$), whereas their arrangement in La$_2$CuO$_4$ is almost flat. The ligand bond angle is known to be the main control parameter for superexchange in cuprates~\cite{khomskii}. Compressing this angle suppresses antiferromagnetic superexchange $J_1$ by reducing the hopping between the same-symmetry half-filled orbitals of Cu$^{2+}$. Concurrently, it facilitates hoppings between $d$-orbitals of different symmetries, which are indeed strongly mixed in the case of CuF$_2$ (Fig.~\ref{fig:dosband}). Such hoppings produce $|{\bf D}|$ in cuprates according to Moriya's mechanism~\cite{mazurenko2008} and give rise to the largely increased $|{\bf D}|/J_1\simeq 0.3$ in CuF$_2$ compared to $|{\bf D}|/J_1\leq 0.01$ in La$_2$CuO$_4$~\cite{wang2023}. 

The DM exchange in La$_2$CuO$_4$ and other parent compounds of high-$T_c$ cuprates is usually treated as a minor term that fixes spin direction and sometimes causes weak ferromagnetism~\cite{shekhtman1992,shekhtman1993}, while having little influence on spin dynamics~\cite{coldea2001}. With its much larger $|{\bf D}|/J_1$, CuF$_2$ may be an interesting platform for studying the effect of DM anisotropy on the quantum spin dynamics of a 2D spin-$\frac12$ antiferromagnet. Our spin-wave calculations for the realistic spin model set the stage for future experimental studies of magnetic excitations in this material. 

In the context of altermagnetism, the case of CuF$_2$ highlights two important limitations of identifying altermagnets from symmetry reasons only. First, symmetry alone does not guarantee that the material shows band splitting and other features of an altermagnet. A ${\bf k}\neq 0$ magnetic order can suppress the splitting, which is the case in CuF$_2$ where interlayer couplings largely differ from those in difluorides with the tetragonal rutile structure. A thorough analysis of magnetic interactions is essential for deciding whether a given material is likely to be an altermagnet. Second, symmetry conditions for altermagnetism often coincide with the symmetry conditions for anisotropic DM interactions~\cite{kimel2024,solovyev2026}. Such interactions are rooted in spin-orbit coupling, which is usually quite weak in $3d$ metals. However, even this weak coupling can have a pronounced effect on the magnetism, as in the case of CuF$_2$ where DM becomes the subleading term. The resulting DM interactions split magnon bands in a manner which is somewhat similar to the band splitting of altermagnets, yet this splitting has an entirely different, relativistic origin.

\section{METHODS}

The magnetic susceptibility of CuF$_2$ was measured using the MPMS SQUID magnetometer from Quantum Design on a commercial CuF$_2$ powder from Alfa Aesar. The hygroscopic CuF$_2$ powder was handled in an Ar-filled glove box, loaded into a tightly closed gelatin capsule, and transferred to the MPMS without contact with air. The sample purity was controlled by x-ray diffraction performed using the Huber G670 camera (CuK$_{\alpha 1}$ radiation, ImagePlate detector). The refined lattice parameters of this sample ($a=3.2926(4)$\,\r A, $b=4.5487(1)$\,\r A, $c=5.9633(1)$\,\r A, $\beta=129.945(1)^{\circ}$, $P2_1/c$) are in good agreement with the literature values~\cite{taylor1974}. No peaks of foreign phases were detected. 

The DFT calculations were done using the generalized gradient approximation (GGA) implemented in the full-potential local-orbital code FPLO version 22~\cite{FPLO}. As a structural input, we used the experimental data from Ref.~\cite{cuf2:fischer74}. The nonmagnetic GGA calculations were performed on a $k$-mesh of $17\times11\times11$ points; the bands crossing the Fermi level were described using an effective one-orbital model by performing Wannier projections~\cite{FPLO_WF} on Cu-centered $d$ states of $x^2\!-\!y^2$ symmetry. The total-energy scalar-relativistic GGA+$U$ calculations were performed using a supercell doubled along $b$ and $c$ axes ($1\times2\times2$ supercell), on a $k$-mesh of $10\times3\times3$ points; we used the Coulomb repulsion $U_d\!=\!8\pm1$\,eV, the Hund exchange $J_d\!=\!1$\,eV, and the fully localized limit as the double-counting correction. For $U_d\!=\!8$\,eV we additionally performed calculations for a nonupled ($1\times3\times3$) supercell on a $k$-mesh of $10\times2\times2$ points. Exchange interactions were estimated by mapping GGA+$U$ total energies onto a classical $S\!=\!\frac12$ Heisenberg model and solving the resulting redundant linear problem by a bootstrapped least-squares procedure.

The exchange anisotropy of $J_1$ was determined using the four-cell method~\cite{DFT_4cell} with GGA+$U$ total energies of noncollinear (relativistic) magnetic configurations calculated on a $2\times2\times2$ $k$-mesh by VASP version 6.4.2~\cite{VASP, VASP_2} with default pseudopotentials~\cite{VASP_pseudo}. All structural relaxations were also performed in VASP.

The uniform and staggered susceptibility of the nearest-neighbor $S=\frac12$ Heisenberg model on the square-lattice was simulated using the quantum Monte Carlo (QMC) loop algorithm implemented in ALPS libraries version 2.3.0. We considered a finite lattice of $N\!=\!24\times24$ sites with periodic boundary conditions, and used 500\,000 (50\,000) sweeps for measurement (thermalization). The thermodynamic properties of the nonfrustrated {$J_1$--\jbp--\jbcp}  model were studied on finite lattices of up to $N\!=\!24\times24\times6$ spins, with 400\,000 (40\,000) sweeps for measurement (thermalization). The exact diagonalizations of the anisotropic $J_1$--$\mathbf{D}$ model were performed on $N\!=\!4\times4$ finite lattices using our own C++ code.
\vspace{0.6cm}

\textbf{Data availability.} 
Experimental and numerical data associated with this work are openly available at \href{https://doi.org/10.5281/zenodo.22145753}{https://doi.org/10.5281/zenodo.22145753}.
\medskip

\textbf{Acknowledgments}
AT acknowledges the computing time made available to him on the high-performance computer at the NHR Center of TU Dresden. This center is jointly supported by the Federal Ministry of Education and Research and the state governments participating in the NHR [www.nhr-verein.de/unsere-partner].
OJ has received funding from the European Innovation Council and SMEs Executive Agency (EISMEA) under the European Union's Horizon Europe research and innovation programme under Grant Agreement n$^{\circ}$\ 101258102. Views and opinions expressed are those of the author(s) only and do not necessarily reflect those of the European Union or EISMEA. Neither the EU nor the granting authority can be held responsible for them.
OJ thanks Ulrike Nitzsche for technical assistance.
IR acknowledges the support by the Engineering and Physical Sciences Research Council, Grant No. EP/V038281/1.
\medskip

\textbf{Author contributions.} 
IR and OJ contributed equally to this work. OJ calculated exchange couplings and fitted magnetic susceptibility data. IR performed mean-field and spin-wave calculations. AT acquired the sample and performed magnetization measurements, as well as relaxations for crystal structure under strain. All co-authors jointly designed the project and prepared the manuscript.
\medskip

\textbf{Competing interests.} The authors declare no competing interests.

\onecolumngrid
\clearpage
\twocolumngrid

\makeatletter
\def\@hangfrom#1{\setbox\@tempboxa\hbox{{#1}}\hskip-\wd\@tempboxa\noindent\box\@tempboxa}
\setcounter{secnumdepth}{2} 
\makeatother

\setcounter{section}{0}
\setcounter{figure}{0}
\setcounter{table}{0}
\setcounter{equation}{0}
\renewcommand{\thefigure}{S\arabic{figure}}
\renewcommand{\thetable}{S\arabic{table}}
\renewcommand{\theequation}{S\arabic{equation}}
\renewcommand{\thesection}{\Roman{section}} 

\begin{onecolumngrid}
  \begin{center}
    {\large \bfseries Supplementing material for \\``Magnon band splitting without altermagnetism in CuF$_2$'' \par}
    \vspace{1.5em}

    {\normalsize Ioannis Rousochatzakis, Oleg Janson, and Alexander A. Tsirlin \par}
    \vspace{0.5em}
    {\small \itshape 
      $^{1}$Department of Physics, Loughborough University, Loughborough LE11 3TU, United Kingdom\\
      $^{2}$Institute for Theoretical Solid State Physics, Leibniz IFW Dresden, 01069 Dresden, Germany\\
      $^{3}$Felix Bloch Institute for Solid-State Physics, University of Leipzig, 04103 Leipzig, Germany \par}
  \end{center}
  \vspace{2em}
\end{onecolumngrid}

\begin{twocolumngrid}
This supplementing material includes technical details and auxiliary information for the microscopic analysis of CuF$_2$ presented in the main text.
\section{Structure and symmetries}

CuF$_2$ crystallizes in the monoclinic space group $P2_1/c$ (No 14) with lattice parameters $a=3.309$\,{\AA}, $b=4.569$\,{\AA}, $c=5.362$\,{\AA}, $\alpha=\gamma=90^\circ$, and $\beta=121.11^\circ$. The Cu sites (Wyckoff positions 2a) reside on an inversion center.  
Figure~\ref{fig:TwoDModel} shows a single $bc$-layer of the structure. The CuF$_4$ units have two orientations, which we denote by A and B in Fig.~\ref{fig:TwoDModel}. The space group includes inversion (at each of the Cu sites), a $2_1$ screw axis along $b$ [$\pi$-rotation around $b$ at $c\!=\!-\frac{3}{4}$, followed by translation by $(0,\frac{1}{2},0)$, i.e., altogether $(x,y,z)\!\mapsto\!(-x,y\!+\!\frac{1}{2},-\frac{3}{2}\!-\!z)$], and a $c$-glide plane perpendicular to $b$ [reflection through $ac$-plane at $b\!=\!\frac{1}{4}$ followed by translation by $(0,0,-\frac{1}{2})$, i.e., altogether $(x,y,z)\!\mapsto\!(x,\frac{1}{2}\!-\!y,z\!-\!\frac{1}{2})$]. For the sites labeled by 1-4 in Fig.~\ref{fig:TwoDModel}, the glide plane maps $(1,2)\mapsto (2,3)$ and the screw axis maps $(2,3)\mapsto(4,2)$.

\begin{figure}[!b]
\includegraphics[width=0.49\textwidth,angle=0,clip=true,trim=0 0 0 0]{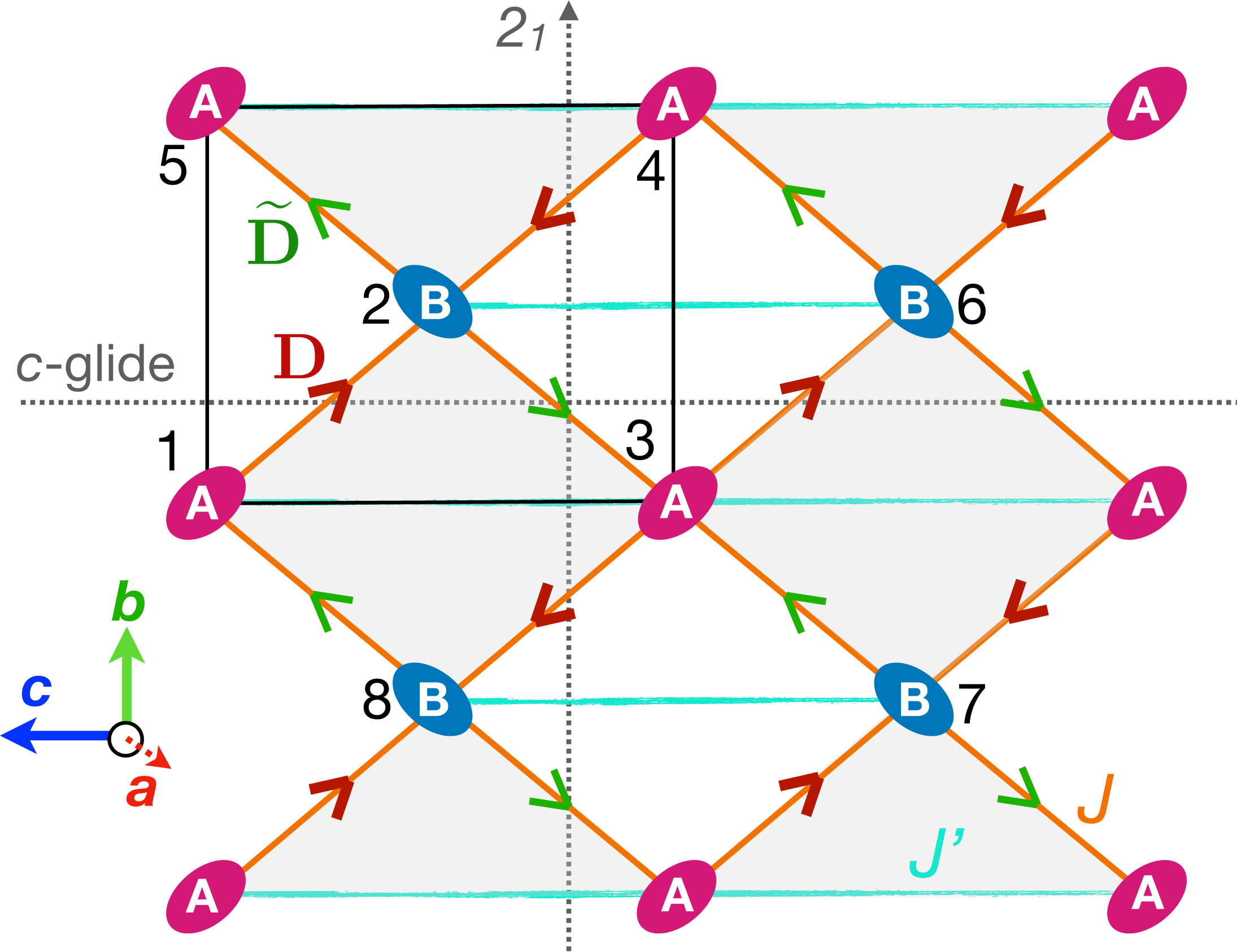}
\caption{\label{fig:TwoDModel} Topology of exchange couplings ($J_1$ and $J_2'$)  and DM vectors (${\bf D}$ and $\widetilde{\bf D}$) in a single $bc$ layer of CuF$_2$. The $c$-glide and the $2_1$ screw axis are also indicated.}
\end{figure}

\section{Microscopic model}
As discussed in the main text, our minimal microscopic model takes the form
\be\label{eq:H1}
\mc{H}=\sum_{i<j}\,\, [J_{ij}\, {\bf S}_i\cdot {\bf S}_j  + {\bf D}_{ij} \cdot {\bf S}_i \times {\bf S}_j]\,,
\ee
where ${\bf S}_i$ and ${\bf S}_j$ denote the spin-1/2 operators for the Cu$^{2+}$ spins at sites $i$ and $j$, and $J_{ij}$ and ${\bf D}_{ij}$ stand, respectively, for the exchange coupling and DM vector on the $(i,j)$ bond.
The dominant exchange coupling is between NN sites, which we denote by $J_1$. There is also a second-neighbour coupling on the diagonals along the $c$ axis, which we denote by $J_2'$  (no such diagonal coupling exists along the $b$ axis). There are also two types of interlayer couplings, which we denote by \jbp\ and \jbcp\ (see main text). 
The DM vectors in a single $bc$ layer are shown in Fig.~\ref{fig:TwoDModel}. The orientations of the DM vectors remain the same in all layers, as adjacent layers are connected by primitive translation along ${\bf a}$. 
There are two, symmetry-related NN DM vectors, ${\bf D}_{12}$ (denoted by ${\bf D}$ in the main text) and ${\bf D}_{25}$ (denoted by $\widetilde{{\bf D}}$ in the following). 
Denoting by $(x,y,z)$ the $(a^\ast, b,c)$ frame, we have
\be
{\bf D}\!=\!(D_{x}, D_{y}, D_{z})\,,~~
\widetilde{{\bf D}}\!\equiv\!(-D_{x}, D_{y}, -D_{z})\,.
\ee

We can re-write Eq.~(\ref{eq:H1}) more explicitly as
\bea\label{eq:H2}
\mc{H} \!&\!=\!&\! 
\sum_{{\bf r}_A}\Big\{  
J_1 {\bf S}_{{\bf r}_A} \cdot \left({\bf S}_{{\bf r}_B}\!+\!{\bf S}_{{\bf r}_B-{\bf b}}\!+\!{\bf S}_{{\bf r}_B+{\bf c}}\!+\!{\bf S}_{{\bf r}_B-{\bf b}+{\bf c}}\right)
\nonumber\\
&&+
J_2' \left(
    {\bf S}_{{\bf r}_A} \cdot {\bf S}_{{\bf r}_A+{\bf c}}\!+\!{\bf S}_{{\bf r}_B} \cdot {\bf S}_{{\bf r}_B+{\bf c}}
    \right)
\nonumber\\
&&+ 
\jbcp\ 
{\bf S}_{{\bf r}_A} \cdot \left({\bf S}_{{\bf r}_A+{\bf a}+{\bf b}+{\bf c}} \!+\! {\bf S}_{{\bf r}_A-{\bf a}-{\bf b}-{\bf c}}\right) 
\nonumber\\
&&+ 
\jbcp\  {\bf S}_{{\bf r}_B} \cdot\left({\bf S}_{{\bf r}_B+{\bf a}-{\bf b}+{\bf c}} \!+\! {\bf S}_{{\bf r}_B-{\bf a}+{\bf b}-{\bf c}}\right)\!
\nonumber\\
&&+ 
\jbp\
{\bf S}_{{\bf r}_A} \cdot ({\bf S}_{{\bf r}_A+{\bf a}+{\bf b}} \!+\! {\bf S}_{{\bf r}_A-{\bf a}-{\bf b}}) \nonumber\\ 
&&+ 
\jbp\ {\bf S}_{{\bf r}_B} \cdot ({\bf S}_{{\bf r}_B+{\bf a}-{\bf b}} \!+\! {\bf S}_{{\bf r}_B-{\bf a}+{\bf b}})
\nonumber\\
&&+
{\bf D}\cdot{\bf S}_{{\bf r}_A} \times \left({\bf S}_{{\bf r}_B}\!+\!{\bf S}_{{\bf r}_B-{\bf b}+{\bf c}}\right)
\nonumber\\
&&-
\widetilde{\bf D} \cdot {\bf S}_{{\bf r}_A} \times \left({\bf S}_{{\bf r}_B-{\bf b}}\!+\!{\bf S}_{{\bf r}_B+{\bf c}}\right)
\Big\} \,,
\eea
where ${\bf r}_A=(n_1,n_2,n_3)$ are the positions of the A sites and ${\bf r}_B={\bf r}_A+(0,\frac{1}{2},\frac{1}{2})$ the positions of the B sites.

According to DFT, the full billinear exchange matrix $\bs{\Lambda}$ for the bond (1,2) of Fig.~\ref{fig:TwoDModel} is, in units of Kelvin, 
\be
\bs{\Lambda} = \left(\begin{array}{ccc}
119.9 &  26.2 & 26.6 \\
-27.0 & 120.8 & 17.2 \\
-25.1 & -13.7 & 121.6
\end{array}\right)\,,
\ee
from which we  obtain the NN Heisenberg exchange 
\be
J_1 = \frac{1}{3}\text{Tr}[\bs{\Lambda}] = 120.8\,\text{K}\,,
\ee
the components of the DM vector ${\bf D}$, $D_\alpha = \epsilon_{\alpha\beta\gamma} (\Lambda^{\beta\gamma}-\Lambda^{\gamma\beta})/2$, 
\be
{\bf D} = (15.5, -25.9, 26.6)\,\text{K}\,,
\ee
and $|{\bf D}|/J=0.33$. 
The symmetric part of the exchange anisotropy is very small 
\be
\bs{\Sigma} \equiv \frac{1}{2} (\bs{\Lambda}+\bs{\Lambda}^T) - J_1 {\bf 1}
= \left(\begin{array}{ccc}
-0.87 &  -0.4 & 0.75 \\
-0.4 & 0.03 & 1.75 \\
0.75 & 1.75 & 0.83
\end{array}\right)\,,
\ee
and can therefore be disregarded.

\section{Classical ground state}

\subsection{Single layer}

The classical ground state of a single layer of CuF$_2$ is a two-sublattice state, with the two spin directions denoted by ${\bf A}$ and ${\bf B}$ (on A and B sites, respectively), as shown in Fig.~\ref{fig:TwoDModel}. 
The spins lie in the plane perpendicular to the vector
\be
{\bf d} \equiv \frac{1}{2}({\bf D}-\widetilde{\bf D}) = (D_{x}, 0, D_{z})\,,
\ee
and are free to rotate globally around this axis. The reason behind this can be seen by looking at the energy contribution from one elementary square of the structure (shaded in Fig.~\ref{fig:TwoDModel}):
\bea\label{eq:Esquare1}
E_{\text{square}} =4 \left(J_1 {\bf A}\cdot {\bf B} + {\bf d}\cdot {\bf A}\times {\bf B}\right)\,,
\eea
where we have disregarded the constant terms arising from the $J_2'$ coupling. So, indeed, the plane of the spins is set by ${\bf d}$ and a global rotation around that vector does not alter the energy at the mean field level.

More explicitly, we can parametrize the general form of the family of degenerate classical ground states in terms of two angles, $\alpha$ and $\phi$, which correspond, respectively, to the global rotation around ${\bf d}$ and the deviation from the ideal antiparallel configuration of ${\bf A}$ and ${\bf B}$ (i.e., the relative angle between ${\bf A}$ and ${\bf B}$ is $\pi\pm \phi$). 
Defining
\be
\hat{\bf d} \!=\! {\bf d}/d\,,~~
\hat{\bf d}_\perp \!=\! \hat{\bf d} \times \hat{\bf y}\!=\!\frac{1}{d} (-D_z,0,D_x)\,,
\ee
where $d=|\bs{d}|=\sqrt{D_x^2+D_z^2}$, the classical ground states take the form
\be\label{eq:2DGS}
\begin{array}{l}
{\bf A}/S = \cos(\alpha+\pi-\frac{\phi}{2})~\hat{\bf y} + \sin(\alpha+\pi-\frac{\phi}{2})~\hat{\bf d}_\perp\,,\\
{\bf B}/S = \cos(\alpha+\frac{\phi}{2})~\hat{\bf y}+\sin(\alpha+\frac{\phi}{2})~\hat{\bf d}_\perp\,.
\end{array}
\ee
We can also describe these states in terms of the total moment ${\bf F}$ and the N\'eel vector ${\bf L}$ (per site, in units of $g\mu_B$),  which are defined by
\be
{\bf F}=\frac{{\bf A}+{\bf B}}{2}\,,~~~
{\bf L}=\frac{{\bf A}-{\bf B}}{2}\,,
\ee
and which satisfy the `spin-length constrains' ${\bf F}\cdot{\bf L}\!=\!0$ and ${\bf F}^2\!+\!{\bf L}^2\!=\!S^2$. 

Given that both the $2_1$ screw axis and the $c$-glide operation map 
\be 
\renewcommand{\arraystretch}{1.3}
\begin{array}{c}
(A_x, A_y, A_z) \mapsto (-B_x, B_y, -B_z)\,,\\
(B_x, B_y, B_z) \mapsto (-A_x, A_y, -A_z)\,,
\end{array}
\ee
it follows that
\be 
\renewcommand{\arraystretch}{1.3}
\begin{array}{c}
(F_x, F_y, F_z) \mapsto (-F_x, F_y, -F_z)\,,\\
(L_x, L_y, L_z) \mapsto (L_x, -L_y, L_z)\,,
\end{array}
\ee
i.e., $L_x$, $F_y$ and $L_z$ transform according to the $\Gamma_1$ irrep (they are even under both $2_1$ and the $c$-glide operation), whereas $F_x$, $L_y$ and $F_z$ transform as the $\Gamma_3$ irrep (they are odd under both $2_1$ and the $c$-glide). 

In terms of ${\bf F}$ and ${\bf L}$, Eq.~(\ref{eq:Esquare1}) becomes
\bea\label{eq:Esquare2}
E_{\text{square}}/4 &=& J_1 ({\bf F}^2-{\bf L}^2) - 2 {\bf d}\cdot {\bf F} \times {\bf L}\,,
\eea
and the parametrization of (\ref{eq:2DGS}) takes the form
\be
\begin{array}{l}
{\bf F} = S~\sin\frac{\phi}{2} \left(-\sin\alpha~\hat{\bf y}+ \cos\alpha~\hat{\bf d}_\perp\right)\,,\\
{\bf L} = S~\cos\frac{\phi}{2} \left(-\cos\alpha~\hat{\bf y} - \sin\alpha~\hat{\bf d}_\perp\right)\,.
\end{array}
\ee
So, in general, the ground states mix both $\Gamma_1$ and $\Gamma_3$.

To proceed we replace (\ref{eq:2DGS}) into (\ref{eq:Esquare1}) and obtain
\be
E_{\text{square}} = - 4S^2 \left( J \cos\phi  + d \sin\phi\right) \,,
\ee
which is independent of $\alpha$, as mentioned above.  Minimizing with respect to $\phi$ gives
\be
\tan\phi = d/J_1\,.
\ee
According to the DFT values given above, $d/J_1\!\simeq\!0.25$, which corresponds to a canting of $\phi\!\simeq\!14^\circ$  away from the fully anti-parallel configuration (i.e., the angle between ${\bf A}$ and ${\bf B}$ is $194^\circ$ or $166^\circ$), with 
\be
|{\bf F}| = 0.062\,,~~
|{\bf L}| =  0.496\,.
\ee
Hence the total moment per site  (for a single layer)  is very small despite the fact that $d$ is $1/4$ of $J_1$.

\subsection{Including the interlayer couplings}
As mentioned above, the CuF$_4$ units have two orientations, A and B. There are two interlayer antiferromagnetic exchange couplings, \jbp\ and \jbcp,  which connect A sites with A sites and B sites with B sites, see Eq.~(\ref{eq:H2}). There is no coupling between A sites in one layer with B sites in another layer. Therefore, taking two layers, 1 and 2, and denoting by ${\bf A}_j$ and ${\bf B}_j$ ($j=1,2$) the A sites and B sites in the two layers,  the energy per unit cell is given by 
\be
\renewcommand{\arraystretch}{1.3}
\begin{array}{l}
E_{\text{layer1}} = 4 \left( J_1 {\bf A}_1 \cdot {\bf B}_1 + {\bf d}\cdot {\bf A}_1 \times {\bf B}_1\right) , 
\\
E_{\text{layer2}} = 4\left(J_1 {\bf A}_2 \cdot{\bf B}_2 + {\bf d}\cdot{\bf A}_2\times{\bf B}_2\right) , 
\\
E_{\text{inter}} = J_{\perp} ({\bf A}_1\cdot{\bf A}_2 + {\bf B}_1\cdot{\bf B}_2),
\end{array}
\ee
where 
\be 
J_\perp \equiv \jbcp\ + \jbp \,,
\ee 
and we have disregarded the constant terms arising from the $J_2'$ couplings. With $J_\perp > 0$, the interlayer exchange can be minimized by 
\be
{\bf A}_2 = -{\bf A}_1\,,~~~
{\bf B}_2 = -{\bf B}_1\,,
\ee
and the angles between A and B sites in all layers are the same. Namely, the interlayer couplings do not frustrate the intra-layer magnetic structure. In the resulting configuration, the FM moment in each layer alternates from one layer to the next, leading to a vanishing total magnetic moment.

\begin{figure}[!t]
\includegraphics[width=0.49\textwidth,angle=0,clip=true,trim=0 0 0 0]{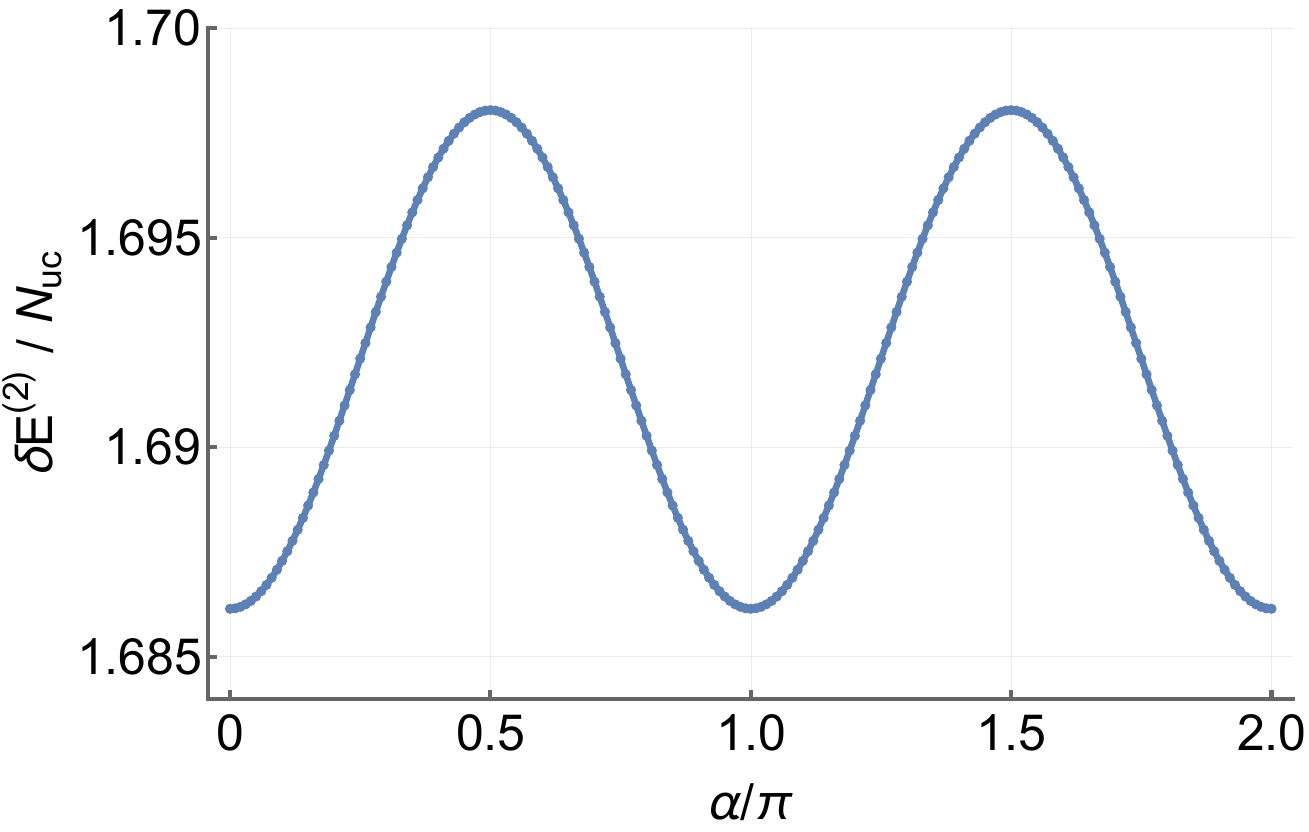}
\caption{\label{fig:ObD2D} Order by disorder effect: Zero-point energy (per unit cell) for a single layer of CuF$_2$, as obtained from a quadratic spin-wave expansion around the classical ground state of (\ref{eq:2DGS}) parametrized by the angle $\alpha$.
}
\end{figure}

\section{State selected by QM fluctuations: \\Order by disorder analysis}

As mentioned above, the global direction of the spins in the plane perpendicular to ${\bf d}$ is not fixed at the mean-field level. This degeneracy is accidental -- the Hamiltonian does not have a U(1) rotation symmetry around ${\bf d}$ -- and is therefore expected to be lifted by quantum fluctuations. To find the preferred orientation of ${\bf L}$ and ${\bf F}$ we carry out a standard quadratric spin-wave expansion and calculate the zero-point energy as a function of the orientation in the plane perpendicular to ${\bf d}$. 

It is sufficient to keep the dominant intra-layer interactions only, so we return to the single layer problem. 
The spin-wave expansion for a single layer with DM interactions included is discussed in Sec.~\ref{sec:LSW2DwithDM} below. We have two sites in the magnetic unit cell and thus there are two magnon branches ($\nu\!=\!1$, $2$), whose dispersions at the quadratic level are given by Eq.~(\ref{eq:omega2}) below and are shown in Fig.~\ref{fig:2DSqw}\,(b) for a representative value of $\alpha$.  The quadratic correction to the ground state energy is given by (see Eq.~(\ref{eq:dE2}) below)
\be
\delta{E}^{(2)} \!=\! \frac{1}{2}\sum_{\nu, {\bf k}} \omega_\nu({\bf k})
\!=\!
\frac{N_{\text{uc}}}{8\pi^2}\!\sum_\nu\!\! \int_{-\frac{\pi}{b}}^{\frac{\pi}{b}}\!\!\!\!dk_y\!\! \int_{-\frac{\pi}{c}}^{\frac{\pi}{c}}\!\!\!\!dk_z~\omega_\nu({\bf k})\,,
\ee
where $N_{\text{uc}}$ is the number of unit cells. 

Figure~\ref{fig:ObD2D} shows the resulting behavior of $\delta{E}^{(2)}$ as a function of $\alpha$, for the coupling parameters found from DFT and given above.  
The energy has minima at $\alpha\!=\!0$ and $\pi$, and maxima at $\alpha\!=\!\pm\pi/2$. The minima of $\delta{E}^{(2)}$ correspond to 
\be
\begin{array}{l}
{\bf A}=S~(0.108, ~0.992, -0.063)\,,\\
{\bf B}=S~(0.108, -0.992, -0.063)\,.
\end{array}
\ee
and
\be
{\bf L} = \pm~0.496~\hat{\bf y}\,,~~~
{\bf F} = \mp~0.062~\hat{\bf d}_\perp\,, 
\ee
where $\hat{\bf d}_\perp\!=\!(-0.864, 0, 0.503)$ in the $xyz$ frame.  
Thus, quantum fluctuations select the $b$ axis for the dominant N\'eel order parameter ${\bf L}$. The corresponding long-range ordering is, therefore, associated with the $\Gamma_3$ irrep, as mentioned in the main text.

\section{Magnetic susceptibility in the PM phase: Mean-field analysis}

In this section, we derive an analytical mean-field expression for the uniform magnetic susceptibility of CuF$_2$ in its PM phase. 
We shall first consider a single layer of CuF$_2$ (disregarding $J_\perp$ and $J_2'$) and carry out a mean-field decoupling of the DM interactions (Sec.~\ref{sec:MF2D}).
We shall then consider the full 3D system and include the couplings $J_\perp$ and $J_2'$ in the mean-field decoupling (Sec.~\ref{sec:MF3D}).

\subsection{Single layer of CuF$_2$}\label{sec:MF2D}

The DM interactions can be decoupled as
\bea
{\bf S}_i \times {\bf S}_j \!&\!\to\!&\! \langle{\bf S}_i\rangle \times {\bf S}_j +{\bf S}_i \times \langle{\bf S}_j\rangle -\langle{\bf S}_i\rangle \times \langle{\bf S}_j\rangle\,,
\eea
and therefore, disregarding overall constants,
\bea
\!\!{\bf D}_{ij} \cdot {\bf S}_i \times {\bf S}_j \!&\!\to\!&\! 
- {\bf h}_{\text{DM},i} \cdot {\bf S}_i - {\bf h}_{\text{DM},j} \cdot {\bf S}_j\,,
\eea
where 
\be
{\bf h}_{\text{DM},i} = {\bf D}_{ij} \times \langle {\bf S}_j \rangle\,,~~
{\bf h}_{\text{DM},j} = {\bf D}_{ji} \times \langle {\bf S}_i \rangle  \,.
\ee
For the two-sublattice state of Fig.~\ref{fig:TwoDModel} and for the sites 2 and 3 we have:
\be
\!\!\!\!
\renewcommand{\arraystretch}{1.4}
\begin{array}{l}
{\bf h}_{\text{DM},A} 
\!\equiv\! 
{\bf h}_{\text{DM},3}
\!=\! \sum_{j=2,6,7,8} {\bf D}_{3j}\times \langle {\bf S}_j\rangle\!=\!+\bs{\delta}\times {\bf B}\,,
\\
{\bf h}_{\text{DM},B} 
\!\equiv\! 
{\bf h}_{\text{DM},2} 
\!=\! \sum_{j=1,3,4,5} {\bf D}_{2j}\times \langle {\bf S}_j\rangle\!=\! -\bs{\delta} \times {\bf A}\,,
\end{array}
\ee
where $\bs{\delta}\!\equiv\!4\bs{d}$. 
The mean-field Hamiltonian takes the form
\be
\mc{H}_{\text{MF}} = \mc{H}_0 - {\bf h}_A \cdot \sum_{i\in A} {\bf S}_i - {\bf h}_B \cdot \sum_{i\in B} {\bf S}_i\,,
\ee
where $\mc{H}_0$ contains the isotropic Heisenberg couplings, ${\bf h}$ is the applied magnetic field divided by $g\mu_B$, and 
\be
\renewcommand{\arraystretch}{1.4}
\!\!\!\begin{array}{l}
{\bf h}_A={\bf h}+{\bf h}_{\text{DM},A}={\bf h}+\bs{\delta}\times {\bf B}\,,\\
{\bf h}_B={\bf h}+{\bf h}_{\text{DM},B}={\bf h}-\bs{\delta}\times {\bf A}\,.
\end{array}
\ee
We can re-write $\mc{H}_{\text{MF}}$ in the form
\be
\mc{H}_{\text{MF}} = \mc{H}_0 - {\bf h}_u \cdot {\bf S}_{\text{tot}} - {\bf h}_s \cdot {\bf L}_{\text{tot}}
\ee 
where
\be
{\bf S}_{\text{tot}} = \sum_{i\in A} {\bf S}_i +\sum_{i\in B} {\bf S}_i\,,~~~
{\bf L}_{\text{tot}} = \sum_{i\in A} {\bf S}_i -\sum_{i\in B} {\bf S}_i\,,
\ee
and 
\be
{\bf h}_u \equiv \frac{1}{2}({\bf h}_A+{\bf h}_B)\,,~~~
{\bf h}_s \equiv \frac{1}{2}({\bf h}_A-{\bf h}_B)\,.
\ee
Let us denote by $\chi_{0,u}$ and $\chi_{0,s}$ the uniform and staggered susceptibilities per site associated with $\mc{H}_0$. 
According to linear response theory, we have:
\be\label{eq:LR1}
{\bf A} = \chi_{0,u} {\bf h}_u + \chi_{0,s} {\bf h}_s\,,~~
{\bf B} = \chi_{0,u} {\bf h}_u - \chi_{0,s} {\bf h}_s \,.
\ee
Now, in the paramagnetic phase, all symmetries must be respected. The symmetries depend on the direction of the external field. 

i) For ${\bf h}\parallel \hat{\bf y}$, the glide plane and the screw axis operations are symmetries also in the presence of the field. These symmetries map 
\be
(A_x,A_y,A_z) \to (-B_x,B_y,-B_z)\,,
\ee
and therefore we must impose $(A_x,A_y,A_z) = (-B_x,B_y,-B_z)$. 
That is, a magnetic field along $y$ induces a uniform magnetization along $y$ and a staggered one in the $xz$ plane. 
The uniform and staggered fields take the form:
\be
\renewcommand{\arraystretch}{1.4}
\!\!\!\begin{array}{l}
{\bf h}_u = (0, h + A_z \delta_x - A_x \delta_z, 0 )\,,~~~\\
{\bf h}_s =  ( - A_y \delta_z, 0 ,  A_y \delta_x)\,.
\end{array}
\ee
According to Eq.~(\ref{eq:LR1}), we have:
\bea
\renewcommand{\arraystretch}{1.4}
\begin{array}{l}
A_x = -\chi_{0,s} A_y \delta_z \,,\\
A_y =  \chi_{0,u} (h+ A_z \delta_x - A_x \delta_z)\,,\\
A_z = \chi_{0,s} A_y \delta_x  \,.
\end{array}
\eea
Solving these equations gives
\be
A_x = -\frac{\bar{\chi}^2 \delta_z}{1-(\delta \bar{\chi})^2}~h\,,~~~
A_z= \frac{\bar{\chi}^2 \delta_x}{1-(\delta \bar{\chi})^2}~h\,,
\ee
and 
\be
A_y= \frac{\chi_{0,u}}{1-(\delta \bar{\chi})^2}~ h\,,
\ee
where
\be
\bar{\chi} \equiv \sqrt{\chi_{0,u}\chi_{0,s}}
\ee
is the `geometric mean' of the uniform and staggered susceptibilities associated with $\mc{H}_0$. 
We therefore find
\be
A_y = B_y =  \chi_{u,yy} h\,,
\ee
where
\be
\chi_{u,yy} =\frac{\chi_{0,u}}{1-(\bar{\chi}\delta)^2}
\ee
is the uniform susceptibility associated with $\mc{H}_{\text{MF}}$ for a field applied along $y$.

ii) For ${\bf h}\perp \hat{\bf y}$,  the glide plane and the screw axis operations are {\bf not} symmetries in the presence of the field. However, if we augment these operations with time-reversal, then we have a symmetry. The combined operations map 
\be
(A_x,A_y,A_z) \to (B_x,-B_y,B_z)\,,
\ee
and, therefore, we must impose $(A_x,A_y,A_z) = (B_x,-B_y,B_z)$. 
That is, a magnetic field in the $xz$ plane induces a uniform magnetization in the $xz$ plane and a staggered one along $y$. 
The uniform and staggered fields now take the form:
\be
\renewcommand{\arraystretch}{1.4}
\!\!\!\begin{array}{l}
{\bf h}_u \!=\! (h_x+A_y \delta_z, 0, h_z-A_y \delta_x)\,,~~~\\
{\bf h}_s \!=\!  (0, -A_z \delta_x+A_x \delta_z, 0)\,.~~~
\end{array}
\ee
Linear response theory then gives
\bea
\renewcommand{\arraystretch}{1.4}
\begin{array}{l}
A_x = \chi_{0,u} (h_x+A_y \delta_z) \,,\\
A_y =  \chi_{0,s}(-A_z \delta_x+A_x \delta_z) \,,\\
A_z = \chi_{0,u} (h_z-A_y \delta_x)\,.
\end{array}
\eea
Solving these equations gives
\be
\renewcommand{\arraystretch}{1.4}
\!\!\!\begin{array}{l}
A_x = \chi_{0,u} \frac{(1- (\bar{\chi}\delta_x)^2) h_x- \delta_x\delta_z \bar{\chi}^2  h_z}{1-(\bar{\chi}\delta)^2}\,,\\
A_y= \frac{\bar{\chi}^2 (\delta_z h_x - \delta_x h_z)}{1-(\bar{\chi}\delta)^2}\,,\\
A_z = \chi_{0,u} \frac{(1- (\bar{\chi}\delta_z)^2) h_z- \delta_x\delta_z \bar{\chi}^2  h_x}{1-(\bar{\chi}\delta)^2}\,.
\end{array}
\ee
From these expressions we find
\be\label{eq:chis2D}
\renewcommand{\arraystretch}{2}
\!\!\!\begin{array}{l}
\chi_{u,xx} = \chi_{0,u}\frac{1-(\bar{\chi}\delta_x)^2}{1-(\bar{\chi}\delta)^2}\,,\\
\chi_{u,zz} = \chi_{0,u}\frac{1-(\bar{\chi}\delta_z)^2}{1-(\bar{\chi}\delta)^2}\,,\\
\chi_{u,xz} =\chi_{u,zx}  = -\chi_{0,u}\frac{\bar{\chi}^2  \delta_x \delta_z}{1-(\bar{\chi}\delta)^2}\,.
\end{array}
\ee

iii) For a general direction of the field, the response is described by the susceptibility tensor
\be
\bs{\chi} = \left(\begin{array}{ccc}
\chi_{u,xx} & 0 & \chi_{u,xz}\\
0 & \chi_{u,yy} & 0 \\
\chi_{u,xz} & 0 & \chi_{u,zz}
\end{array}\right)\,.
\ee
Note that the average over all three directions gives 
\be\label{eq:chiavg2D}
\chi_{\text{avg}}\!\equiv\! \frac{1}{3}\left(\chi_{u,xx} \!+\! \chi_{u,yy} \!+\! \chi_{u,zz} \right) \!=\! \chi_{0,u}\frac{1- \frac{1}{3}(\bar{\chi}\delta)^2}{1-(\bar{\chi}\delta)^2}.~~~
\ee

In the isotropic limit, $\delta\to 0$, we recover $\chi_{xx}=\chi_{yy}=\chi_{zz}=\chi_{0,u}$, $\chi_{xz}=0$, and $\chi_{\text{avg}}=\chi_{0,u}$.

\subsection{3D system: Including $J_\perp$ and $J_2'$}\label{sec:MF3D}

If we decouple the exchange couplings $J_2'$ and $J_\perp$ (besides the DM couplings) we obtain the following contribution to the local fields at the A and B sites:
\be
\renewcommand{\arraystretch}{1.4}
\!\!\!\begin{array}{l}
{\bf h}_{A} \!=\! {\bf h}-2 J_p {\bf A}+\bs{\delta}\times {\bf B} \,,\\
 {\bf h}_{B} \!=\! {\bf h}- 2 J_p {\bf B}-\bs{\delta}\times {\bf A}\,,
\end{array}
\ee
where 
\be 
J_p = J_2'+J_{\perp}\,,
\ee 
and we used the fact that we are in the PM phase and, therefore, all A sites are parallel to each other and the same is true for the B sites. 
We proceed as before and consider fields along $y$-axis first and then fields in the $xz$-plane. 
\\

i) For ${\bf h}\parallel \hat{\bf y}$, we have again:
\be
(B_x,B_y,B_z) \to (-A_x,A_y,-A_z)\,.
\ee
The uniform and staggered fields take the form:
\be
\renewcommand{\arraystretch}{1.4}
\!\!\!\begin{array}{l}
{\bf h}_u = (0, h -2 J_p A_y + A_z \delta_x - A_x \delta_z, 0 )\,,~~~\\
{\bf h}_s =  (-2J_p A_x - A_y \delta_z, 0 , -2J_p A_z + A_y \delta_x)\,.
\end{array}
\ee
Linear response theory then gives
\bea
\renewcommand{\arraystretch}{1.45}
\begin{array}{l}
A_x = -\chi_{0,s} (A_y \delta_z + 2J_p A_x) \,,\\
A_y =  \chi_{0,u} (h-2 J_p A_y  + A_z \delta_x - A_x \delta_z)\,,\\
A_z = \chi_{0,s} (A_y \delta_x - 2J_p A_z) \,.
\end{array}
\eea
Solving these equations gives
\be
\renewcommand{\arraystretch}{1.4}
\!\!\!\begin{array}{l}
A_x = -\bar{\chi}^2~ \delta_z~h/f\,,\\
A_y= \chi_{0,u} (1+2J_p\chi_{0,s})~ h/f\,,\\
A_z= \bar{\chi}^2~ \delta_x~h/f\,,
\end{array}
\ee
where
\bea
\!\!
f \!=\!(1\!+\!2J_p\chi_{0,u})(1\!+\!2J_p\chi_{0,s})\!-\!(\delta\bar{\chi})^2\,.~~~
\eea
From these expressions we find
\be
\chi_{u,yy} =\chi_{0,u} (1+2J_p\chi_{0,s})/f\,.
\ee

ii) For ${\bf h}\perp \hat{\bf y}$,  we have, for a given layer, as before:
\be
(B_x,B_y,B_z) \to (A_x,-A_y,A_z)\,.
\ee
The uniform and staggered fields now take the form:
\be
\renewcommand{\arraystretch}{1.4}
\!\!\!\begin{array}{l}
{\bf h}_u \!=\! (h_x\!-\!2J_p A_x\!+\!A_y \delta_z, 0, h_z\!-\!2J_p A_z\!-\! A_y \delta_x)\,,~~~\\
{\bf h}_s \!=\!  (0, -2J_p A_y \!-\!A_z \delta_x\!+\!A_x \delta_z, 0)\,.~~~
\end{array}
\ee
Linear response theory then gives
\bea
\renewcommand{\arraystretch}{1.4}
\begin{array}{l}
A_x = \chi_{0,u} (h_x-2J_p A_x+A_y \delta_z) \,,\\
A_y =  \chi_{0,s}(-2J_p A_y -A_z \delta_x+A_x \delta_z) \,,\\
A_z = \chi_{0,u} (h_z-2J_p A_z-A_y \delta_x)\,.
\end{array}
\eea
Solving these equations gives
\be
\renewcommand{\arraystretch}{1.4}
\!\!\!\begin{array}{l}
A_x = \chi_{0,u} \frac{(f+ (\bar{\chi}\delta_z)^2) h_x- \delta_x\delta_z \bar{\chi}^2  h_z}{(1+2J_p \chi_{0,u})~f}\,,\\
A_y= \bar{\chi}^2 (\delta_z h_x - \delta_x h_z)/f\,,\\
A_z = \chi_{0,u} \frac{(f+ (\bar{\chi}\delta_x)^2) h_z- \delta_x\delta_z \bar{\chi}^2  h_x}{(1+2J_p \chi_{0,u})~f}\,.
\end{array}
\ee
From these expressions we find
\be\label{eq:chis3D}
\renewcommand{\arraystretch}{2}
\!\!\!\begin{array}{l}
\chi_{u,xx} = \frac{ \chi_{0,u} (f+(\bar{\chi}\delta_z)^2)}{(1+2J_p \chi_{0,u})~f}\,,\\
\chi_{u,zz} = \frac{\chi_{0,u} (f+(\bar{\chi}\delta_x)^2)}{(1+2J_p \chi_{0,u})~f}\,,\\
\chi_{u,xz} = -\frac{\chi_{0,u} \bar{\chi}^2  \delta_x \delta_z}{(1+2J_p \chi_{0,u})~f}\,.
\end{array}
\ee
Finally, the average over all three directions gives 
\be\label{eq:chiavg3D}
\chi_{\text{avg}}\!=\! \chi_{0,u}\frac{f + \frac{2}{3}(\bar{\chi}\delta)^2}{(1+2J_p \chi_{0,u})~f}\,.
\ee
In the absence of $J_p$, these expressions reduce to the ones we derived previously [Eqs.~(\ref{eq:chis2D}) and (\ref{eq:chiavg2D})], whereas in the absence of DM interactions (i.e., for $\delta=0$) we obtain:
\be\label{eq:chiavg3Ddelta0}
\chi_{\text{avg}}(\delta=0)\!=\!\frac{\chi_{0,u}}{1+2J_p\, \chi_{0,u}}\,.
\ee
Equation~(\ref{eq:chiavg3D}) is the expression used for the magnetic susceptibility fit of CuF$_2$ (see main text).

\section{Spin-wave theory}

In the following, we discuss the spin-wave dispersions in a number of different cases. We shall first outline the main steps of the linear spin wave theory and introduce the basic concepts and quantities of interest in Sec.~\ref{sec:LSWgeneral}. In Sec.~\ref{sec:LSW2DwithoutDM}, we discuss the case of a single layer of CuF$_2$ without DM couplings. 
In Sec.~\ref{sec:LSW2DwithDM}, we include the DM couplings and discuss the magnon splitting of a single layer arising from the DM couplings.
We then move on to the 3D case  and consider first a scenario where the interlayer couplings \jbcp\ and \jbp\ are ferromagnetic and not antiferromagnetic (Sec.~\ref{sec:LSW3DwithoutDMJperpFM}). This scenario is, therefore, not relevant for CuF$_2$, but it is worth discussing it because the corresponding magnetic order is altermagnetic (it resembles the situation in MnF$_2$~\cite{Bandyopadhyay2025}), and we wish to contrast the resulting spin-wave splitting with the splitting in CuF$_2$ (Secs.~\ref{sec:LSW3DwithoutDMJperpAM} and \ref{sec:LSW3DwithDMJperpFM}), which is not altermagnetic in origin.

\subsection{General definitions}
\label{sec:LSWgeneral}
We begin by outlining the main steps of linear spin wave expansion and the key quantities of interest. 
Suppose our reference classical ground state around which we wish to expand has a magnetic unit cell with $p$ sites. We can then label the positions of the spins by $({\bf R},\mu)$, where ${\bf R}$ is the position of the magnetic unit cell and $\mu$ is the sublattice index. Furthermore, for each sublattice, we can define a local, orthonormal, right-handed axis frame $\{{\bf u}_\mu, {\bf v}_\mu, {\bf w}_\mu\}$, where ${\bf w}_\mu$ is the direction of the spins at $({\bf R},\mu)$ in the reference classical state. One can then express the spin operator ${\bf S}_{{\bf R},\mu}$ in the corresponding local frame as 
\be
{\bf S}_{{\bf R},\mu} = S^u_{{\bf R},\mu} {\bf u}_\mu
+S^v_{{\bf R},\mu} {\bf v}_\mu
+S^w_{{\bf R},\mu} {\bf w}_\mu\,,
\ee
and then perform a Holstein-Primakoff transformation~\cite{HP1940}, keeping only the leading terms, 
\be
\renewcommand{\arraystretch}{1.4}
\begin{array}{l}
S^w_{{\bf R},\mu} = S-a^+_{{\bf R},\mu}a_{{\bf R},\mu}\,,
\\
S^+_{{\bf R},\mu} \equiv S^u_{{\bf R},\mu}+ i S^v_{{\bf R},\mu}
\approx \sqrt{2S} a_{{\bf R},\mu}\,,
\\
S^-_{{\bf R},\mu} \equiv S^u_{{\bf R},\mu} - i S^v_{{\bf R},\mu}
\approx \sqrt{2S} a^+_{{\bf R},\mu}\,,
\end{array}
\ee
where $a_{{\bf R},\mu}$ and $a^+_{{\bf R},\mu}$ are the bosonic operators that quantify the spin fluctuations at the site $({\bf R},\mu)$. Replacing into the spin Hamiltonian and carrying out a Fourier transformation (using $a_{{\bf k},\mu}\!=\!\frac{1}{\sqrt{N_{uc}}}\sum_{\bf k} e^{i {\bf k}\cdot {\bf R}} a_{{\bf R},\mu}$, where $N_{uc}$ is the number of magnetic unit cells) leads to the following quadratic bosonic form
\bea\label{eq:HLSW1}
\mc{H}^{(2)} \!&\!=\!&\! E_{\text{cl}}/S \!+\! \frac{1}{2} \sum_{{\bf k},\mu\nu}
\Big[
C_{\mu\nu} a^\dagger_{{\bf k},\mu} a_{{\bf k},\nu}
\!+\!F_{\mu\nu} a^\dagger_{{\bf k},\mu} a^\dagger_{-{\bf k},\nu} \nonumber\\
&&+D_{\mu\nu} a_{-{\bf k},\mu} a_{{\bf k},\nu}+G_{\mu\nu} a_{-{\bf k},\mu} a^\dagger_{-{\bf k},\nu}\Big]\,,
\eea
or, in matrix form, 
\be\label{eq:HLSW2}
\mc{H}^{(2)} = E_{\text{cl}}/S + \frac{1}{2} \sum_{\bf k} \vec{A}_{\bf k}^+ \!\cdot\! \vec{M}_{\bf k} \!\cdot\! \vec{A}_{\bf k}
\ee
where $E_{\text{cl}}$ is the classical energy, 
\bea
\vec{A}_{\bf k}^+ &=& \left(
\begin{array}{c}
a^\dagger_{{\bf k},1}, \cdots, a^\dagger_{{\bf k},p},  a_{-{\bf k},1}, \cdots, a_{-{\bf k},p}
\end{array}
\right)\,,
\eea
and
\be\label{eq:Mk}
\renewcommand{\arraystretch}{1.4}
\vec{M}_{\bf k} = \left(\begin{array}{c|c}
\vec{C}_{\bf k} & \vec{F}_{\bf k} \\
\hline
\vec{F}^\ast_{-{\bf k}} & \vec{C}^\ast_{-{\bf k}}
\end{array}\right)\,,
\ee
where ${\bf C}_{\bf k}$ and ${\bf F}_{\bf k}$ are $p\times p$ matrices. 

To diagonalize $\mc{H}^{(2)}$, we perform a standard Bogoliubov transformation~\cite{Blaizot} ${\bf A}_k = {\bf U}_k \!\cdot\! {\bf B}_k$, where ${\bf B}_k$ contains the new set of bosonic operators
\be
{\bf B}_{\bf k}^+ = \left(
\begin{array}{c}
b^\dagger_{{\bf k},1}, \cdots, b^\dagger_{{\bf k},p},  b_{-{\bf k},1}, \cdots, b_{-{\bf k},p}
\end{array}
\right)\,,
\ee
and the matrix ${\bf U}_{\bf k}$ is such that $\bs{\Omega}_{\bf k}\equiv{\bf U}_{\bf k}^+ \cdot {\bf M}_{\bf k} \cdot {\bf U}_{\bf k}$ becomes diagonal. The transformation must also preserve the bosonic commutation relations, which can be expressed compactly as ${\bf g}= {\bf U}_{\bf k} \cdot {\bf g}\cdot  {\bf U}_{\bf k}^+$, where ${\bf g}$ is the ``commutator'' matrix ${\bf g}=\text{diag}({\bf 1},-{\bf 1})$, and ${\bf 1}$ is the $p\times p$ identity matrix. The above two conditions give 
\be
({\bf g} {\bf M}_{\bf k})\cdot {\bf U}_{\bf k} = {\bf U}_{\bf k} \cdot \left( {\bf g} \bs{\Omega}_{\bf k} \right)\,,
\ee
which is an eigenvalue equation (the columns of ${\bf U}_{\bf k}$ contain the eigenvectors of ${\bf g} {\bf M}_{\bf k}$). One can further show~\cite{Blaizot} that if ${\bf M}_{\bf k}$ is semi-definite positive then ${\bf g}\bs{\Omega}_{\bf k} \!=\! \text{diag}\left(\omega_{1}({\bf k}), \cdots, \omega_{p}({\bf k}), -\omega_{1}({\bf k}), \cdots, -\omega_{p}({\bf k})\right)$ and
\be
\mc{H}^{(2)}=E_{\text{cl}}/S +\sum_{{\bf k},\mu}\Big( b_{{\bf k},\mu}^+ b_{{\bf k},\mu} +\frac{1}{2} \Big) \omega_\mu({\bf k})\,.
\ee

The ground state is the vacuum of the $b$ bosons. Therefore, the quadratic correction to the ground-state energy is given by the standard expression
\be\label{eq:dE2}
\delta E^{(2)} = E_{\text{cl}}/S+ \sum_{{\bf k},\mu} \frac12 \omega_\mu({\bf k})\,.
\ee
The order by disorder effect discussed earlier (see Fig.~\ref{fig:ObD2D}) is based on this expression.

Using the Bogoliubov transformantion 
\be 
a_{{\bf k}, \mu} = \sum\nolimits_{\nu} (U_{{\bf k},\mu\nu} b_{{\bf k},\nu}+U_{{\bf k},\mu, p+\nu} b_{-{\bf k},\nu}^+)\,,
\ee
one can also obtain an expression for the zero-temperature local spin length (denoted by $\langle S_i\rangle$ in the main text): 
\be
\langle S^w_{{\bf R},\mu} \rangle = S - \frac{1}{N_{uc}} \sum_{{\bf k},\nu} |U_{\mu, p+\nu}({\bf k})|^2\,.
\ee
The spin length results shown in Fig.~6 of the main text are based on this expression.

Finally, the one-magnon contribution to the dynamical spin structure factor components are given by 
\be
\!\!\!
\mc{S}^{\alpha\beta}({\bf q},\omega) = \frac{N_{uc}S}{2}  \!\sum\nolimits_{\mu}\! X_{\mu}^\alpha({\bf q}) X_{\mu}^{\beta\ast}({\bf q}) \delta(\omega- \omega_\mu({\bf q})),~~
\ee
where 
\bea
{\bf X}_{\mu}({\bf q}) \!&\!=\!&\! \sum\nolimits_{\nu}\!e^{i {\bf q}\cdot\bs{\rho}_\nu} f_\nu({\bf q})~\delta{\bf m}_{\nu}^{(\mu)}({\bf q}),\label{eq:Xq}
\\
\delta{\bf m}_{\nu}^{(\mu)}({\bf q}) \!&\!=\!&\! U_{\nu,\mu}({\bf q})~{\bf e}_{\nu}^{-} \!+\! U_{p+\nu,\mu}({\bf q})~{\bf e}_{\nu}^{+},\label{eq:deltamuq}
\eea
$\alpha,\beta=x,y,z$, $\bs{\rho}_\nu$ is the physical position of the $\nu$-th ion inside the magnetic unit cell, $f_\nu({\bf q})$ is its magnetic form factor, and ${\bf e}_\nu^\pm\equiv{\bf u}_\nu\pm i {\bf v}_\nu$. The quantity $\delta{\bf m}_{\nu}^{(\mu)}({\bf q})$ represents the FT of the magnetization fluctuations (in units of $g\mu_B=1$) of the $\nu$-th site, for the given spin-wave mode $\mu$.

The dynamical structure factor component measured in unpolarized  neutron scattering experiments is given by
\be
\mc{S}_\perp({\bf q},\omega) = 
\frac{N_{uc}S}{2}  \!\sum\nolimits_{\mu}\! |{\bf X}_{\mu\perp}({\bf q})|^2~\delta(\omega- \omega_\mu({\bf q})),
\ee
where ${\bf X}_{\mu\perp}\equiv\hat{\bf q}\times({\bf X}_\mu \times \hat{\bf q})$. The results shown in Fig.~7 of the main text and in Fig.~\ref{fig:2DSqw} below are based on this expression.

Finally, the so-called `chiral factor' that can be measured in polarized neutron scattering experiments with neutron polarization ${\bf P}\parallel{\bf q}$ is given by 
\be\label{eq:Cqw}
\mc{C}({\bf q},\omega)\!=\!-i\frac{N_{uc}S}{2} \!\sum_{\mu}\! \hat{\bf q}\cdot \big({\bf X}_{\mu}({\bf q})\times 
{\bf X}^\ast_{\mu}({\bf q})\big) 
 \delta(\omega- \omega_\mu({\bf q})).
\ee

\begin{figure*}[!t]
\includegraphics[width=0.99\textwidth,angle=0,clip=true,trim=0 0 0 0]{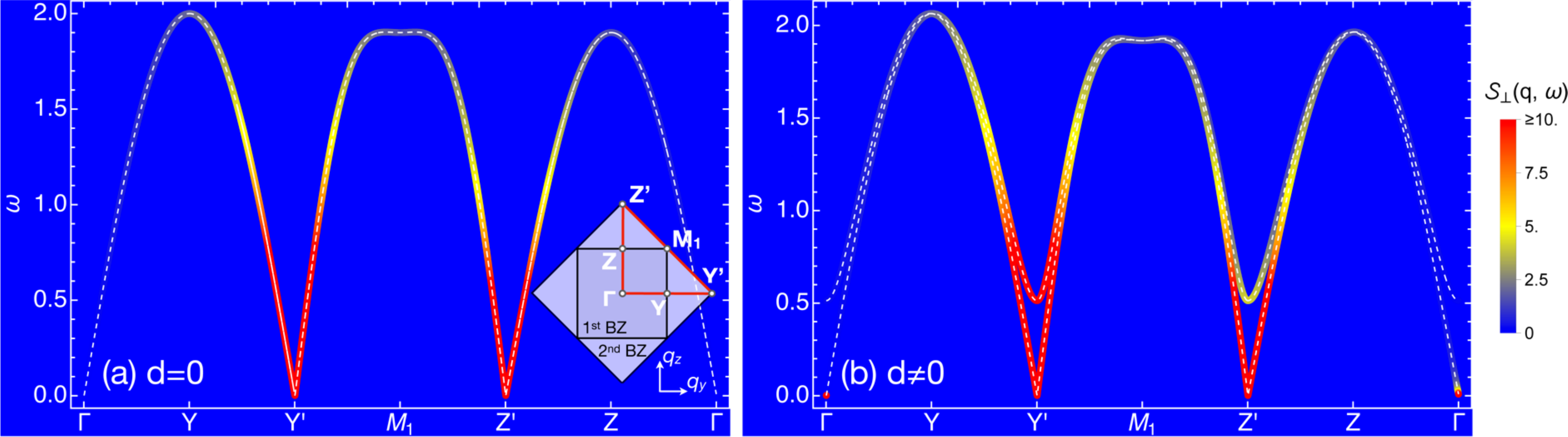}
\caption{Linear spin-wave theory results for a single $bc$-layer of CuF$_2$, described by the $J_1$-$J_2'$-DM model (see text) with $J_1=121$\,K, $J_2'=6$\,K, and with $d=0$ (panel a) and $d\neq 0$ (panel b). Magnon dispersions (in units of $J_1$) along special directions in the $bc$ scattering plane (see inset) are shown by dashed lines. Superimposed (coloured) symbols show the unpolarized neutron scattering intensity $\mc{S}_\perp({\bf q},\omega)$. For (b), the classical state around which we expand is one of the states selected by quantum fluctuations.}
\label{fig:2DSqw}
\end{figure*}

\begin{figure*}[!t]
\includegraphics[width=0.99\textwidth,angle=0,clip=true,trim=0 0 0 0]{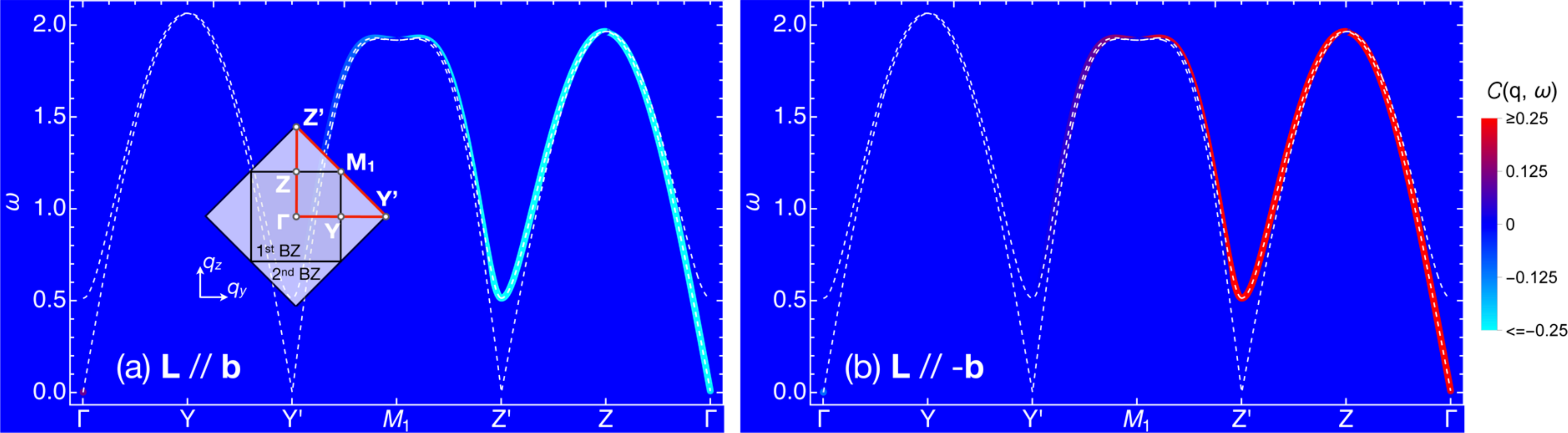}
\caption{Chiral factor $\mc{C}({\bf q},\omega)$ for the two magnon branches of Fig.~\ref{fig:2DSqw}\,(b), which correspond to a single layer of CuF$_2$ described by the $J_1$-$J_2'$-DM model. The two panels correspond to N\'eel vector parallel (a) and antiparallel (b) to the $b$ axis.}
\label{fig:2DCqw}
\end{figure*}

\subsection{Single layer} 
\subsubsection{Single layer without DM couplings}\label{sec:LSW2DwithoutDM}

We consider a single $bc$-layer of CuF$_2$. We have a collinear phase, with two spin sublattices ($p=2$): 
\be\label{eq:sites2D} 
\hspace{-2cm}
\renewcommand{\arraystretch}{1.4}
\begin{array}{ll}
\text{\textbullet}~
{\bf r}=(0,0,0) &\text{(A site; spin up)}
\\
\text{\textbullet}~
{\bf r}=(0,\frac{1}{2},\frac{1}{2}) & \text{(B site; spin down)}
\end{array}
\ee
Here, the $2\times2$ matrices ${\bf C}_{\bf k}$ and ${\bf F}_{\bf k}$ are given by
\be\label{eq:CFDG1}
{\bf C}_{\bf k} 
\!=\!\!\left(\!\!\begin{array}{cc}\!
\alpha_{\bf k} & 0 \\
0 & \alpha_{\bf k} 
\end{array}\!\!\right),
~~~
{\bf F}_{\bf k}\!=\!\!\left(\!\!\begin{array}{cc}\!
0 & e^{i p_{\bf k}}\beta_{\bf k} \\
e^{-i p_{\bf k}} \beta_{\bf k}  & 0 
\end{array}\!\!\right)\,,
\ee
where
\be\label{eq:abp}
\renewcommand{\arraystretch}{1.4}
\begin{array}{l}
\alpha_{\bf k} =  2J_1 + J_2'(\cos{\kappa_c}-1)\,,\\
\beta_{\bf k} = 2J_1 \cos(\frac{\kappa_b}{2})\cos(\frac{\kappa_c}{2}) \,, 
\\
p_{\bf k}=(\kappa_b-\kappa_c)/2\,,
\end{array}
\ee
and $\kappa_b\!=\!{\bf k}\cdot{\bf b}\!=\!k_y b$ and $\kappa_c\!=\!{\bf k}\cdot{\bf c}\!=\!k_z c$. 
Diagonalizing ${\bf g} {\bf M}_{\bf k}$ gives the magnon frequencies 
\be\label{eq:omega1}
\omega_{1}({\bf k}) = \omega_{2}({\bf k}) = \sqrt{\alpha_{\bf k}^2-\beta_{\bf k}^2}\,.
\ee

\subsubsection{Single layer with DM couplings}\label{sec:LSW2DwithDM}

Let us now include the DM interactions in the single layer problem discussed above. 
Unlike the Heisenberg terms, the DM interactions know about the difference between A and B sites, and therefore we expect two split magnon branches. 

The quadratic spin-wave Hamiltonian takes again the general form of (\ref{eq:HLSW2}) with 
\be\label{eq:CFDG2}
\!\!{\bf C}_{\bf k} \!=\! \left(\!\!\begin{array}{cc}\!
\alpha_{\bf k}' & e^{i p_{\bf k}} \gamma_{\bf k} \\
 e^{-i p_{\bf k}} \gamma_{\bf k}^\ast & \alpha_{\bf k}'
\end{array}\!\!\right)\!,~
{\bf F}_{\bf k}\!=\! \left(\!\!\begin{array}{cc}\!
0 & e^{i p_{\bf k}} \beta'_{\bf k} \\
 e^{-i p_{\bf k}} \beta'_{\bf k} & 0 
\end{array}\!\!\right)\!,
\ee
where
\be
\renewcommand{\arraystretch}{1.4}
\begin{array}{l}
\alpha_{\bf k}'=2(J_1\cos\phi+d\sin\phi)+J_2'(\cos{\kappa_z}-1)\,,
\\
\beta_{\bf k}' = \lambda_2 f_2({\bf k})  + 2 i~D_y f_1({\bf k}) \cos\alpha \cos\frac{\phi}{2}\,,
\\
\gamma_{\bf k} = \lambda_1 f_2({\bf k}) + 2 i~D_y f_1({\bf k}) \sin\alpha \sin\frac{\phi}{2}\,,
\end{array}
\ee
and 
\be\label{eq:apbpc}
\renewcommand{\arraystretch}{1.4}
\begin{array}{c}
\lambda_1\!=\!J_1(1\!-\!\cos\phi)\!-\!d\sin\phi\,,
\\
\lambda_2\!=\!J_1(1\!+\!\cos\phi)\!+\!d\sin\phi\,,
\\
f_1({\bf k})\!=\!\sin{\kappa_b}\sin{\kappa_c}\,,
\\
f_2({\bf k})\!=\!\cos\frac{\kappa_b}{2}\cos\frac{\kappa_c}{2}\,.
\end{array}
\ee
The magnon frequencies are given by
\be\label{eq:omega2}
\omega_\pm({\bf k}) = \sqrt{(\alpha_{\bf k}'\pm|\gamma_{\bf k}|)^2-|\beta'_{\bf k}|^2}\,.
\ee
As expected, the expressions (\ref{eq:CFDG2}) and (\ref{eq:omega2}) reduce to (\ref{eq:CFDG1}) and (\ref{eq:omega1}), respectively, when $d\!=\!0$.

Note that, in order to obtain the spin-wave excitations above the states selected by quantum fluctuations, we must replace $\alpha=0$ or $\pi$ to the above expressions. For these states, $\gamma_{\bf k} \propto f_2({\bf k})$, which vanishes for $\kappa_b=\pm\pi$ and or $\kappa_c=\pm\pi$. Hence, the splitting $\omega_+-\omega_-$ vanishes along the edges of the first BZ.

\subsubsection{Results for a single layer}
Figure~\ref{fig:2DSqw} shows the magnon dispersions and corresponding data for $\mc{S}_\perp({\bf q},\omega)$, with and without DM interactions. The results are analogous to those shown in Fig.~7 of the main text for the 3D system.  

In the absence of DM anisotropy (Fig.~\ref{fig:2DSqw}\,(a)), the two magnon branches are degenerate throughout the momentum plane, as in the case of $J_2'\!=\!0$, the standard NN Heisenberg model on the square lattice. The origin of this degeneracy can be understood based on the fact that  a sublattice spin rotation on the down spin (B) sites does not lower the translational symmetry of the spin Hamiltonian (which is larger than the actual one, since the Heisenberg couplings do not differentiate between A and B sites).
Turning to $\mc{S}_\perp({\bf q},\omega)$, the results show the expected Goldstone modes at the magnetic Bragg peak positions $(0,1,0)$ (point Y') and $(0,0,1)$ (point Z').

Figure~\ref{fig:2DSqw}\,(b) shows the corresponding results when DM terms are included. 
The two branches are no longer degenerate (except for the edges of the first BZ, as mentioned above), and their splitting is largest in the region around the $\Gamma$ point. There is one zero-energy mode remaining, but this is an artifact of linear spin-wave theory, and reflects the accidental ground state degeneracy  at the classical level (associated with the choice of the angle $\alpha$ in Eq.~(\ref{eq:2DGS})). This zero-energy mode is then expected to gap out by nonlinear spin-wave corrections.

Turning to $\mc{S}_\perp({\bf q},\omega)$, the main features are as follows: i)  $\mc{S}_\perp({\bf q},\omega)$ is large for the pseudo-Goldstone modes at the magnetic Bragg peak positions;
ii) Compared to panel (a), we now have a weak intensity at the $\Gamma$ point. This reflects the small uniform FM moment generated by the DM anisotropy; 
iii) The intensity of the second mode at $(0,1,0)$ is much larger than that of the second mode at $(0,0,1)$. This asymmetry, which is also present in the 3D system (see main text), reflects the special role of the $b$-axis.

Finally, we discuss our results for the chiral magnon factor $\mc{C}({\bf q},\omega)$ for a single layer of CuF$_2$. In the absence of DM anisotropy, the chiral factor (not shown) vanishes for all ${\bf q}$ and $\omega$. By contrast, as shown in  Fig.~\ref{fig:2DCqw}, for $d\!\neq\!0$, one of the magnon branches acquires a nonzero chirality when $q_z\neq0$ (path $Y'\to M_1\to Z'\to Z\to\Gamma$).  The sign of $\mc{C}({\bf q},\omega)$ depends on the direction of the N\'eel vector in the reference ground state: The chiral factor is negative (positive) for ${\bf L}$ parallel (antiparallel) to the $b$ axis. This stems from the fact that the two reference states ($\alpha\!=\!0$ and $\alpha\!=\!\pi$) are related to each other by time reversal. As a result, going from $\alpha\!=\!0$ to $\pi$, the vectors ${\bf X}_\mu({\bf q})\!\to\!{\bf X}_\mu^\ast({\bf q})$ (module an unimportant numerical phase factor), and therefore, according to (\ref{eq:Cqw}), $\mc{C}({\bf q},\omega)$ changes sign.

\subsection{3D system} 
\subsubsection{3D system with $\jbcp, \jbp <0$ and no DM interactions}\label{sec:LSW3DwithoutDMJperpFM}

\begin{figure}[!t]
\includegraphics[width=0.49\textwidth,angle=0,clip=true,trim=0 0 0 0]{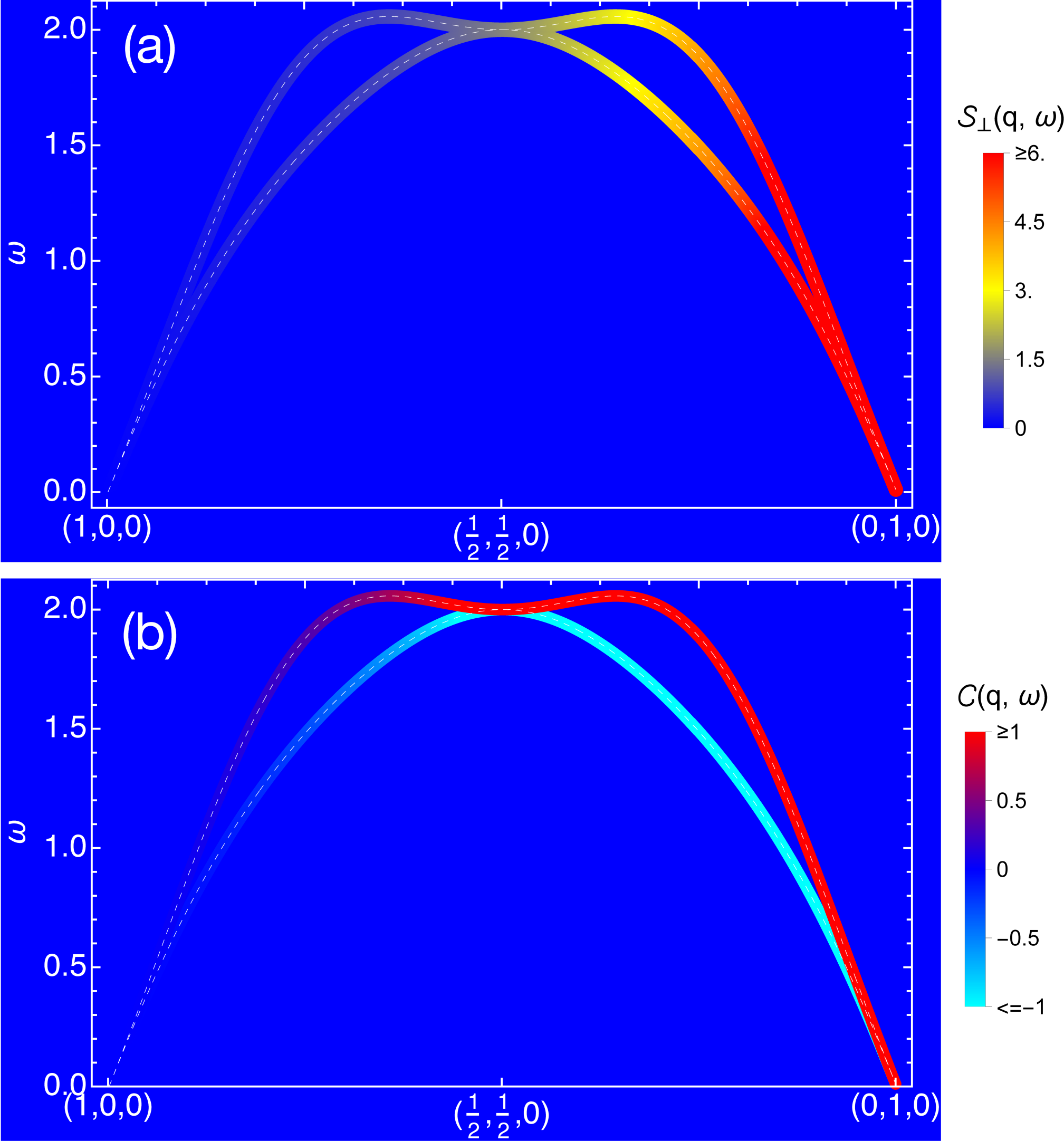}
\caption{Linear spin-wave theory results for the hypothetical altermagnetic scenario (not relevant for CuF$_2$) where the interlayer couplings $J_{bc}^+$ and $J_b^+$ are ferromagnetic, and there are no DM interactions present. Dashed lines show the magnon dispersions (in units of $J_1$) along special directions in momentum space (in r.l.u.).  Superimposed (coloured) symbols show $\mc{S}_\perp({\bf q},\omega)$ (panel a) and $\mc{C}_({\bf q},\omega)$ (panel b).}
\label{fig:3DFMlayersSqwCqw}
\end{figure}

We now switch to the multi-layer, 3D system. As mentioned above, we shall first consider the `altermagnetic' scenario (not relevant for CuF$_2$), where the interlayer couplings \jbcp\ and \jbp\ are ferromagnetic and not antiferromagnetic. For our purposes, we shall disregard the DM interactions. 

With \jbcp\ and \jbp\ both negative, all layers feature the same magnetic structure, the magnetic primitive vectors are ${\bf b}$, ${\bf c}$ and ${\bf a}$, and the magnetic unit cell contains the two sites of (\ref{eq:sites2D}) of the single layer case. The $2\times2$ matrices ${\bf C}_{\bf k}$ and ${\bf F}_{\bf k}$ take the form
\be\label{eq:CFDG3}
\!\!\!\!\!
\renewcommand{\arraystretch}{1.4}
\begin{array}{l}
{\bf C}_{\bf k}
\!=\!\!\left(\!\begin{array}{c@{\hskip -0.07cm}c}\!
\alpha_{\bf k}\!+\!h_{\bf k}& 0 \\
0 & \alpha_{\bf k}\!+\!h_{\bf k}'
\end{array}\!\right)\!,~
{\bf F}_{\bf k}\!=\!\!\left(\!\begin{array}{c@{\hskip -0.07cm}c}\!
0 & e^{i p_{\bf k}}\beta_{\bf k} \\
e^{-i p_{\bf k}} \beta_{\bf k}  & 0 
\end{array}\!\right)\!,
\end{array}
\ee
where $\alpha_{\bf k}$, $\beta_{\bf k}$ and $p_{\bf k}$ are given in (\ref{eq:abp}) above, and
\be
\!\!\!
\renewcommand{\arraystretch}{1.3}
\begin{array}{l}
h_{\bf k}/2 \!=\!{\jbp} \cos(\kappa_a\!+\!\kappa_b)\!+\!{\jbcp} \cos(\kappa_a\!+\!\kappa_b\!+\!\kappa_c)-J_\perp,
\\
h'_{\bf k}/2 \!=\! {\jbp} \cos(\kappa_a\!-\!\kappa_b)\!+\!{\jbcp} \cos(\kappa_a\!-\!\kappa_b\!+\!\kappa_c)-J_\perp,
\end{array}
\ee
$\kappa_a\!=\!{\bf k}\cdot {\bf a}\!=\!a [\cos(\beta-\frac{\pi}{2}) k_x\!-\!\sin(\beta-\frac{\pi}{2}) k_z]$,  and, as before, $\kappa_b\!=\!k_y b$ and $\kappa_c\!=\!k_z c$. 
The Bogoliubov diagonalization gives the following magnon dispersions
\be\label{eq:omega3}
\omega_{\pm}({\bf k}) \!=\! \pm \frac{h'_{\bf k}\!-\!h_{\bf k}}{2}\!+\!\sqrt{\left(\alpha_{\bf k}\!+\!\frac{h_{\bf k}\!+\!h'_{\bf k}}{2}\right)^2\!-\!\beta_{\bf k}^2}\,.
\ee
Hence, we find a nonzero spin-splitting $\Delta({\bf k})\equiv\omega_{+}-\omega_{-}$, which is controlled by {\jbp} and {\jbcp}:
\be
\Delta({\bf k})\!=\! h'_{\bf k}\!-\!h_{\bf k}\!=\!4\sin \kappa_b [{\jbp}  \sin{\kappa_a}\!+\!{\jbcp} \sin(\kappa_a+\kappa_c)].
\ee
This is analogous to the altermagnetic spin splitting in the rutile structure of MnF$_2$~\cite{Bandyopadhyay2025}.

Figure~\ref{fig:3DFMlayersSqwCqw} shows numerical results for the magnon dispersions and corresponding data for $\mc{S}_\perp({\bf q},\omega)$ (panel a) and $\mc{C}({\bf q},\omega)$ (panel b). This figure demonstrates the altermagnetic magnon splitting and the opposite chirality of the two magnons.

\subsubsection{3D system without DM and with $\jbcp, \jbp >0$}\label{sec:LSW3DwithoutDMJperpAM}
 
We now return to the situation in CuF$_2$, where the interlayer couplings are antiferromagnetic. Let us first disregard the DM interactions; we shall include these in the next subsection. The magnetic order is collinear and the unit cell doubles also along the $a$ axis (because of the different orientations of the CuF$_4$ units on A and B sites). Namely the magnetic primitive vectors are ${\bf b}$, ${\bf c}$ and $2{\bf a}$, and the magnetic unit cell contains four sites:
\be\label{eq:sites3D} 
\hspace{-2cm}
\renewcommand{\arraystretch}{1.4}
\begin{array}{ll}
\text{\textbullet}~
{\bf r}=(0,0,0) &\text{(A site; spin up)}
\\
\text{\textbullet}~
{\bf r}=(0,\frac{1}{2},\frac{1}{2}) & \text{(B site; spin down)}
\\
\text{\textbullet}~
{\bf r}=(1,0,0) &\text{(A site; spin down)}
\\
\text{\textbullet}~ 
{\bf r}=(1,\frac{1}{2},\frac{1}{2}) &\text{(B site; spin up)}
\end{array}
\ee
From symmetry grounds, this order is not altermagnetic, as the up spin at $(0,0,0)$ is connected to the down spin at $(1,0,0)$ by a primitive translation. Hence, any spin splitting is not altermagnetic in origin (see below).

The quadratic spin-wave Hamiltonian takes again the general form of (\ref{eq:HLSW2}) where now
\be
\!\!\!
\vec{A}_{\bf k}^+ \!=\!\left(\!\!
\begin{array}{c}
a^\dagger_{{\bf k},1},  \cdots, a^\dagger_{{\bf k},4},  a_{-{\bf k},1}, \cdots, a_{-{\bf k},4}
\end{array}
\!\!\right)\,,
\ee
and the matrices ${\bf C}_{\bf k}$ and ${\bf F}_{\bf k}$ of (\ref{eq:Mk}) are given by
\be
{\bf C}_{\bf k} \!=\!\widetilde{\alpha}_{\bf k}~{\bf 1}_{4} \,,
\ee
and  
\be
{\bf F}_{\bf k}
\!=\!
\left(\!\!\begin{array}{cccc}\!
0 & e^{i p_{\bf k}} \beta_{\bf k} & e^{i \kappa_a} \eta_{1{\bf k}} &0\\
e^{-i p_{\bf k}} \beta_{\bf k}  & 0&0& e^{i \kappa_a} \eta_{2{\bf k}}\\
e^{-i \kappa_a} \eta_{1{\bf k}} & 0&0&e^{i p_{\bf k}} \beta_{\bf k}\\
0 & e^{-i \kappa_a} \eta_{2{\bf k}}&e^{-i p_{\bf k}} \beta_{\bf k}&0
\end{array}\!\!\right)\!,
\ee
where 
\be
\widetilde{\alpha}_{\bf k}\!=\!\alpha_{\bf k}\!+\!2J_\perp\,,
\ee
$(\alpha_{\bf k}, \beta_{\bf k}, p_{\bf k})$ are given in (\ref{eq:abp}) above, and 
\be\label{eq:gamma12}
\!\!\!
\renewcommand{\arraystretch}{1.4}
\begin{array}{l}
\eta_{1{\bf k}}/2 = \jbcp \cos(\kappa_a+\kappa_b+\kappa_c)+\jbp \cos(\kappa_a+\kappa_b)\,,
\\
\eta_{2{\bf k}} /2 = \jbcp \cos(\kappa_a-\kappa_b+\kappa_c)+\jbp \cos(\kappa_a-\kappa_b)\,.
\end{array}
\ee
A Bogoliubov transformation gives four magnon branches, each two-fold degenerate: 
\be\label{eq:omega4}
\renewcommand{\arraystretch}{1.4}
\begin{array}{l}
\omega_{1}({\bf k})\!=\!\omega_{2}({\bf k})  \!=\! \sqrt{\widetilde{\alpha}_{\bf k}^2\!-\!\big[
    \sqrt{\beta_{\bf k}^2\!+\!\eta_{-,{\bf k}}^2}
    \!+\! 
    \eta_{+,{\bf k}}\big]^2},
\\
\omega_{3}({\bf k}) \!=\!\omega_{4}({\bf k}) 
\!=\! 
\sqrt{\widetilde{\alpha}_{\bf k}^2\!-\!\big[
    \sqrt{\beta_{\bf k}^2\!+\!\eta_{-,{\bf k}}^2}\!-\! \eta_{+,{\bf k}}\big]^2},
\end{array}
\ee
where 
\be 
\eta_{\pm,{\bf k}}\equiv(\eta_{1{\bf k}}\pm\eta_{2{\bf k}})/2\,.
\ee 

We should note again here that the Heisenberg model has a larger translational symmetry group than the actual one, since the isotropic couplings do not contain any information about the different orientation of CuF$_4$ units on A and B sites. This means that, for the Heisenberg model, one can use a smaller magnetic unit cell with the two sites of (\ref{eq:sites2D}), and magnetic primitive vectors chosen as ${\bf b}$, ${\bf c}$, and ${\bf a}+{\bf b}$.  
In turn, this means that the two pairs of magnon branches of (\ref{eq:omega4}) are expected to unfold into a one, four-fold degenerate branch in the BZ corresponding to the two-site unit cell. This is also expected on the basis of the fact that a sublattice spin rotation of the down spin sites does not lower the translational symmetry of the Heisenberg model. 
The reason we used the four-site unit cell is to compare with the results of the next subsection, where we include the DM couplings, which know about the difference between A and B sites.

\subsubsection{3D system with DM and $\jbcp, \jbp >0$}
\label{sec:LSW3DwithDMJperpFM}

We finally consider the 3D system with DM interactions included. The magnetic order in each layer is the two-sublattice coplanar order of Sec.~\ref{sec:LSW2DwithDM}, and adjacent layers order antiferromagnetically, so the total FM canting ${\bf F}$ cancels out. 
The magnetic unit cell contains the four sites of (\ref{eq:sites3D}), with primitive vectors ${\bf b}$, ${\bf c}$ and $2{\bf a}$. 

The quadratic spin-wave Hamiltonian takes again the form (\ref{eq:HLSW2}) with
\be 
{\bf C}_{\bf k}\!=\!\left(\!\begin{array}{llll}
 {\widetilde{\alpha}'}_{\bf k} &  e^{i p_{\bf k}} \gamma_{\bf k} &  0 & 0
 \\
 e^{-i p_{\bf k}} \gamma^\ast_{\bf k} & {\widetilde{\alpha}'}_{\bf k} & 0 & 0
 \\
 0 & 0 & {\widetilde{\alpha}'}_{\bf k} & e^{i p_{\bf k}} \gamma^\ast_{\bf k}
 \\ 
 0 & 0 & e^{-i p_{\bf k}} \gamma_{\bf k} &  {\widetilde{\alpha}'}_{\bf k}
 \end{array}\!\right),
\ee
\be 
{\bf F}_{\bf k}\!=\!\left(\!\begin{array}{llll}
0 & e^{i p_{\bf k}} \beta'_{\bf k} & e^{i \kappa_a} \eta_{1{\bf k}} &  0
\\
e^{-i p_{\bf k}} \beta'_{\bf k} & 0 & 0 & e^{i \kappa_a} \eta_{2{\bf k}}
\\
e^{-i \kappa_a} \eta_{1{\bf k}} & 0 & 0 & e^{i p_{\bf k}} {\beta_{\bf k}'}\!^\ast 
\\
0 & e^{-i \kappa_a} \eta_{2{\bf k}} &  e^{-i p_{\bf k}} {\beta_{\bf k}'}\!^\ast & 0
\end{array}\!\right)\,,
\ee 
where 
\be 
{\widetilde{\alpha}'}_{\bf k}\equiv\alpha'_{\bf k}+2J_\perp\,,
\ee
$(\alpha'_{\bf k}, \beta'_{\bf k}, \gamma_{\bf k})$ are given by (\ref{eq:apbpc}), and $(\eta_{1{\bf k}},\eta_{2{\bf k}})$
are given by (\ref{eq:gamma12}). 
The magnon frequencies take the form
\be 
\renewcommand{\arraystretch}{1.4}
\begin{array}{l}
\omega_1({\bf k})=\sqrt{\mc{A}_{\bf k}+2\sqrt{\mc{B}_{\bf k}+\mc{C}_{\bf k}}}\,,
\\
\omega_2({\bf k})=\sqrt{\mc{A}_{\bf k}+2\sqrt{\mc{B}_{\bf k}-\mc{C}_{\bf k}}}\,,
\\
\omega_3({\bf k})=\sqrt{\mc{A}_{\bf k}-2\sqrt{\mc{B}_{\bf k}+\mc{C}_{\bf k}}}\,,
\\
\omega_4({\bf k})=\sqrt{\mc{A}_{\bf k}-2\sqrt{\mc{B}_{\bf k}-\mc{C}_{\bf k}}}\,,
\end{array}
\ee
where
\be 
\renewcommand{\arraystretch}{1.4}
\begin{array}{l}
\mc{A}_{\bf k} = \widetilde{\alpha}_{\bf k}^{\prime 2}+|\gamma_{\bf k}|^2-\eta_{+,{\bf k}}^2-\eta_{-,{\bf k}}^2\,,
\\
\mc{B}_{\bf k}=(\widetilde{\alpha}_{\bf k}^{\prime 2}-\eta_{-,{\bf k}}^2)|\gamma_{\bf k}|^2+2|{\widetilde{\alpha}'}_{\bf k} \beta'_{\bf k} \gamma_{\bf k} \eta_{+,{\bf k}}|\,,
\\  
\mc{C}_{\bf k}=(|\beta'_{\bf k}|^2+\eta_{-,{\bf k}}^2) \eta_{+,{\bf k}}^2\,.
\end{array}
\ee

The magnon dispersions and corresponding unpolarized neutron scattering intensity $\mc{S}_\perp({\bf q},\omega)$ are shown in Fig.~7 of the main text, with and without DM interactions.  
As for the chiral structure factor, $\mc{C}({\bf q},\omega)$, this vanishes for all ${\bf q}$ and $\omega$ (see main text).

\begin{suppbib}{3}
\makeatletter
\providecommand \@ifxundefined [1]{%
 \@ifx{#1\undefined}
}%
\providecommand \@ifnum [1]{%
 \ifnum #1\expandafter \@firstoftwo
 \else \expandafter \@secondoftwo
 \fi
}%
\providecommand \@ifx [1]{%
 \ifx #1\expandafter \@firstoftwo
 \else \expandafter \@secondoftwo
 \fi
}%
\providecommand \natexlab [1]{#1}%
\providecommand \enquote  [1]{``#1''}%
\providecommand \bibnamefont  [1]{#1}%
\providecommand \bibfnamefont [1]{#1}%
\providecommand \citenamefont [1]{#1}%
\providecommand \href@noop [0]{\@secondoftwo}%
\providecommand \href [0]{\begingroup \@sanitize@url \@href}%
\providecommand \@href[1]{\@@startlink{#1}\@@href}%
\providecommand \@@href[1]{\endgroup#1\@@endlink}%
\providecommand \@sanitize@url [0]{\catcode `\\12\catcode `\$12\catcode
  `\&12\catcode `\#12\catcode `\^12\catcode `\_12\catcode `\%12\relax}%
\providecommand \@@startlink[1]{}%
\providecommand \@@endlink[0]{}%
\providecommand \url  [0]{\begingroup\@sanitize@url \@url }%
\providecommand \@url [1]{\endgroup\@href {#1}{\urlprefix }}%
\providecommand \urlprefix  [0]{URL }%
\providecommand \Eprint [0]{\href }%
\providecommand \doibase [0]{https://doi.org/}%
\providecommand \selectlanguage [0]{\@gobble}%
\providecommand \bibinfo  [0]{\@secondoftwo}%
\providecommand \bibfield  [0]{\@secondoftwo}%
\providecommand \translation [1]{[#1]}%
\providecommand \BibitemOpen [0]{}%
\providecommand \bibitemStop [0]{}%
\providecommand \bibitemNoStop [0]{.\EOS\space}%
\providecommand \EOS [0]{\spacefactor3000\relax}%
\providecommand \BibitemShut  [1]{\csname bibitem#1\endcsname}%
\let\auto@bib@innerbib\@empty
\bibitem [{\citenamefont {Bandyopadhyay}\ \emph {et~al.}(2025)\citenamefont
  {Bandyopadhyay}, \citenamefont {Raj}, \citenamefont {Ghosez}, \citenamefont
  {Pujari},\ and\ \citenamefont {Bhowal}}]{Bandyopadhyay2025}%
  \BibitemOpen
  \bibfield  {author} {\bibinfo {author} {\bibfnamefont {S.}~\bibnamefont
  {Bandyopadhyay}}, \bibinfo {author} {\bibfnamefont {A.}~\bibnamefont {Raj}},
  \bibinfo {author} {\bibfnamefont {P.}~\bibnamefont {Ghosez}}, \bibinfo
  {author} {\bibfnamefont {S.}~\bibnamefont {Pujari}},\ and\ \bibinfo {author}
  {\bibfnamefont {S.}~\bibnamefont {Bhowal}},\ }\bibfield  {title} {\bibinfo
  {title} {Rational control of magnonic and electronic band splittings},\
  }\href {https://doi.org/10.1103/c3kh-5s5t} {\bibfield  {journal} {\bibinfo
  {journal} {Phys. Rev. B}\ }\textbf {\bibinfo {volume} {112}},\ \bibinfo
  {pages} {L100402} (\bibinfo {year} {2025})}\BibitemShut {NoStop}%
\bibitem [{\citenamefont {Holstein}\ and\ \citenamefont
  {Primakoff}(1940)}]{HP1940}%
  \BibitemOpen
  \bibfield  {author} {\bibinfo {author} {\bibfnamefont {T.}~\bibnamefont
  {Holstein}}\ and\ \bibinfo {author} {\bibfnamefont {H.}~\bibnamefont
  {Primakoff}},\ }\bibfield  {title} {\bibinfo {title} {Field dependence of the
  intrinsic domain magnetization of a ferromagnet},\ }\href
  {https://doi.org/10.1103/PhysRev.58.1098} {\bibfield  {journal} {\bibinfo
  {journal} {Phys. Rev.}\ }\textbf {\bibinfo {volume} {58}},\ \bibinfo
  {pages} {1098} (\bibinfo {year} {1940})}\BibitemShut {NoStop}%
\bibitem [{\citenamefont {Blaizot}\ and\ \citenamefont
  {Ripka}(1985)}]{Blaizot}%
  \BibitemOpen
  \bibfield  {author} {\bibinfo {author} {\bibfnamefont {J.-P.}\ \bibnamefont
  {Blaizot}}\ and\ \bibinfo {author} {\bibfnamefont {G.}~\bibnamefont
  {Ripka}},\ }\href@noop {} {\emph {\bibinfo {title} {Quantum Theory of Finite
  Systems}}}\ (\bibinfo  {publisher} {MIT Press},\ \bibinfo {address}
  {Cambridge, Massachusetts},\ \bibinfo {year} {1985})\ Chap.~\bibinfo
  {chapter} {3}\BibitemShut {NoStop}%
\end{suppbib}%

\end{twocolumngrid}

\end{document}